\documentclass[aps,showpacs,twocolumn,amsmath,amsfonts,amssymb,longbibliography,prl]{revtex4-1}
\usepackage{graphicx}
\usepackage[tight]{subfigure}
\usepackage{times}
\usepackage{color}
\usepackage{bm}
\usepackage{array}
\usepackage[colorlinks=true,bookmarks=false,citecolor=blue,linkcolor=red,urlcolor=blue]{hyperref}
\usepackage{appendix}

\begin{document}

\title{Spinon Fractionalization from Dynamic Structure Factor of Spin-$1/2$ Heisenberg Antiferromagnet on the Kagome Lattice}
\author {W. Zhu$^1$, Shou-Shu Gong$^{2,3}$, D. N. Sheng$^3$}
\affiliation{$^1$Theoretical Division, T-4 and CNLS, Los Alamos National Laboratory, Los Alamos, New Mexico 87545, USA\\
$^2$Department of Physics, Beihang University, Beijing, 100191, China\\
$^3$Department of Physics and Astronomy, California State University, Northridge, California 91330, USA}

\begin{abstract}
We study dynamical spin structure factor (DSSF) of $S=1/2$ Heisenberg model on the kagome lattice (KAFM) by means of density-matrix renormalization group.
By comparison with the well-defined magnetic ordered state and chiral spin liquid sitting nearby in the phase diagram, the KAFM with the nearest-neighbor interaction shows distinct dynamical response behaviors.
First of all, the DSSF displays important spectral intensity predominantly at low frequency region around $\mathbf Q=M$ point in momentum space, and shows a broad  spectral distribution at high frequency region for momenta along the boundary of the extended Brillouin zone.
Secondly, spinon excitation spectrum is identified from momentum and energy resolved DSSF, which shows critical behavior with much
reduced spectrum intensity comparing to the neighboring chiral spin liquid.
By adding a weak Dzyaloshinkii-Moriya interaction, the DSSF demonstrates a strong sensitivity to the
boundary conditions more consistent with  a gapless spin liquid.
These results capture the main observations in the inelastic neutron scattering measurements of herbertsmithite, and indicate the spin liquid nature of the ground state with fractionalized spinon excitations.
By following the DSSF crossing  the quantum phase transition  between the CSL and the magnetical ordered phase, we identify  the spinon condensation  driving  the quantum phase transition.
\end{abstract}


\maketitle

{\it Introduction.---}
Quantum spin liquid (QSL) is a novel quantum phase which behaves differently from conventional magnetic states~\cite{Leon2010,PALee2006,Leon2017}.
It does not show any magnetic order or lattice symmetry breaking even approaching zero temperature limit.
Theoretical studies have shown the intrinsic nature of QSL including massive entanglement and fractionalized excitations~\cite{Wen1989,WenNiu1990,Wen1991,Read1991}, which are challenging to be measured directly in experiments.
Experimentally, QSL candidates have been identified in frustrated magnetic materials such as kagome- and triangular-lattice compounds~\cite{mendels2007,helton2007,han2012,fu2015,yamashita2008,shimizu2003,kurosaki2005,norman2016}.
The kagome antiferromagnet herbertsmithite~\cite{mendels2007,helton2007,han2012,fu2015} is one of the most promising spin-liquid candidates.
The magnetic order of the material has been excluded down to temperatures a few orders~\cite{helton2007,han2012} of magnitude below coupling energy scale.
Furthermore, the inelastic neutron scattering (INS) measurement characterizes the dynamic spin structure factor (DSSF) $\mathcal{S}({\bf Q}, \omega)$ as a broad continuum spectrum in higher frequency regime~\cite{han2012}, which is distinctly different from the spectrum of conventional magnon excitations.
It remains an open issue what information regarding topological order of the state, and the fractionalization of spin excitations can be extracted from such measurements.
In particular,  some other factors in material such as disorder may also lead to a similar continuum of $\mathcal{S}({\bf Q}, \omega)$~\cite{shimokawa2015}, making theoretical understanding of the INS essential for distinguishing different physics.
In addition, whether such a spin-liquid candidate has a finite spin gap remains unresolved in experimental probes including the INS~\cite{han2012} and the nuclear magnetic resonance (NMR)~\cite{fu2015}.
To clarify these questions, theoretical studies on dynamic measurements related to experimental probes are highly desired.

In theoretical study, the ground state of the spin-$1/2$ kagome antiferromagnet (KAFM) with nearest-neighbor Heisenberg interaction has been investigated intensively~\cite{sachdev1992,waldtmann1998,Ran2007,Hermele2008,Iqbal2011,Iqbal2013,Iqbal2014,Yan2011,Depenbrock2012,Jiang2012nature,Jiang2008,messio2012,lauchli2016,mei2016,jiang2016}, which captures the dominant interaction for the herbertsmithite.
Although a QSL ground state has been established in the KAFM, the full nature of the QSL including the nature of the fractionalized quasi-particles, and the existence of a spin gap is still under debate.
While earlier density matrix renormalization group (DMRG) simulation suggested a gapped spin liquid~\cite{Yan2011,Depenbrock2012,Jiang2012nature,Jiang2008}, parton construction and variational Monte Carlo study found the optimized ground state as the gapless U(1) Dirac spin liquid~\cite{Ran2007,Hermele2008,Iqbal2011,Iqbal2013}.
Such a gapless spin liquid scenario is also indirectly supported by recent DMRG targeting the system response to the inserted flux and tensor network results~\cite{he2017,liao2017}. However, more direct evidence from low energy excitations
are still absent.
The open question regarding the nature of the QSL phase demands new theoretical approaches beyond the ground state study such as the DSSF that reveals excitation properties.
So far most of the studies on the DSSF of the KAFM are based on the mean-field analysis or approximate methods~\cite{messio2010,dodds2013,punk2014,punk2016, sherman2018}, the unbiased numerical model calculation is rare limited to small systems~\cite{lauchli2009}.

\begin{figure*}
\includegraphics[width=0.22\linewidth]{phase_diagram.eps}
\includegraphics[width=0.21\linewidth]{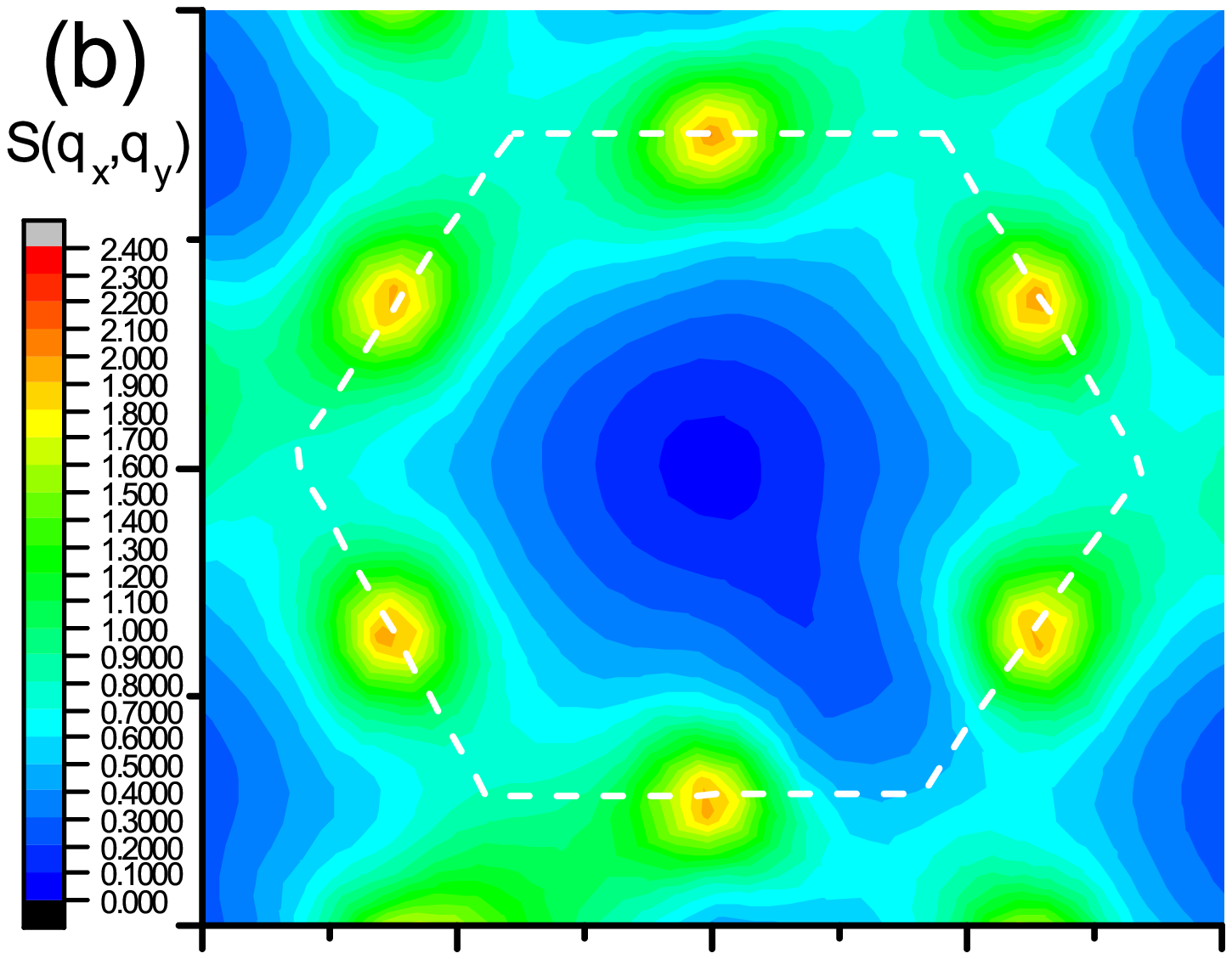}
\includegraphics[width=0.21\linewidth]{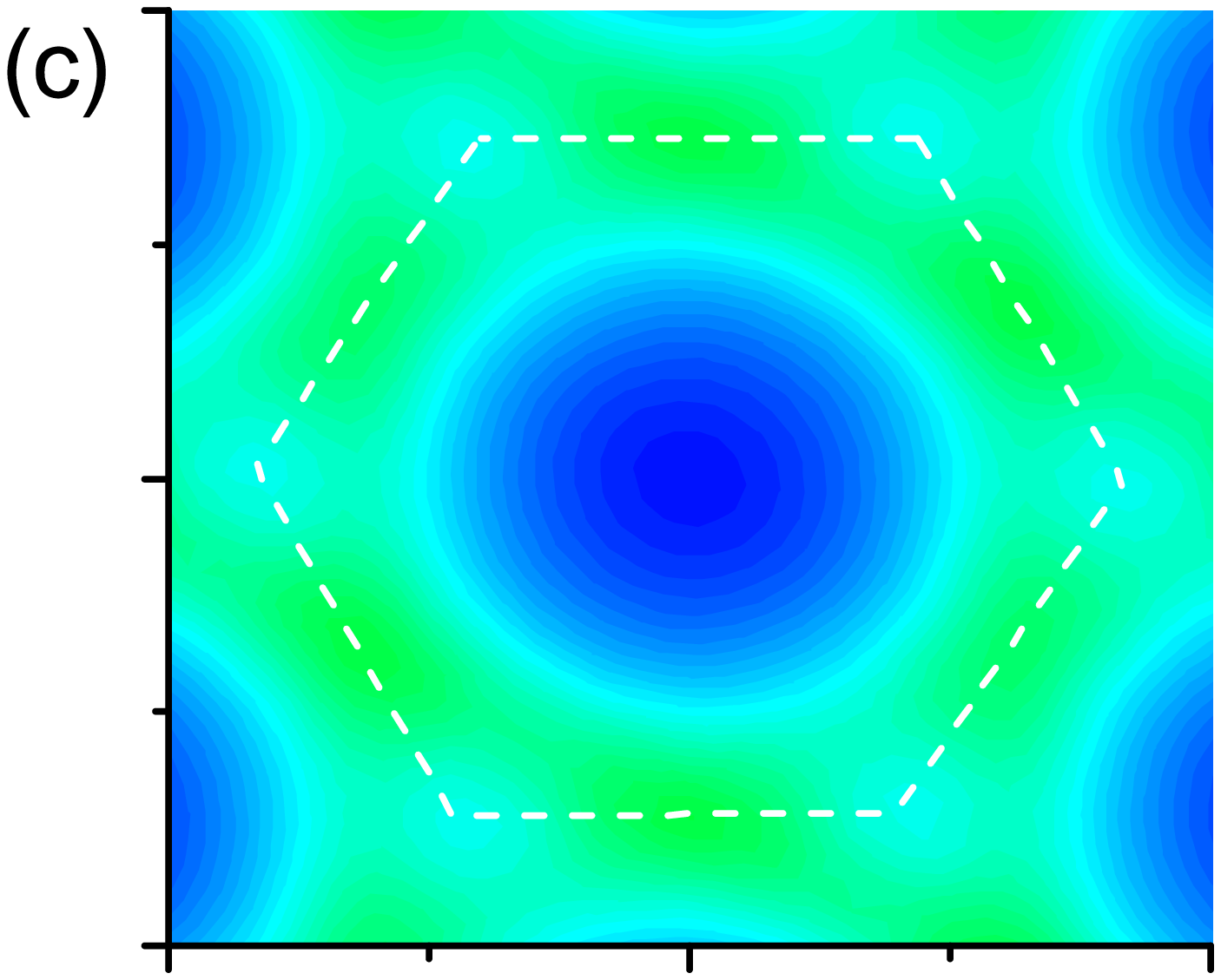}
\includegraphics[width=0.21\linewidth]{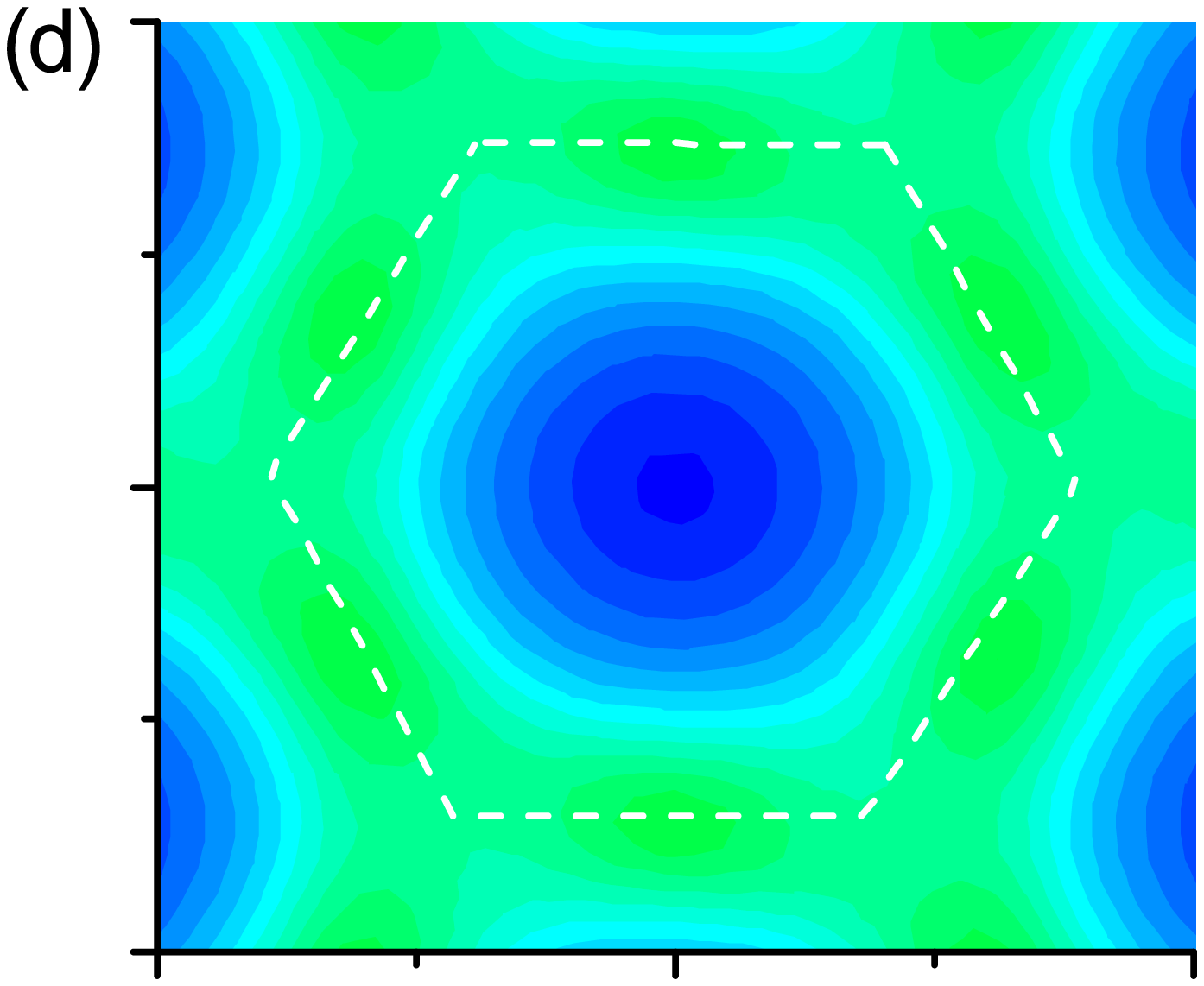}
\caption{Static spin structure factor of the kagome model in different quantum phases.
(a) Quantum phase diagram of the kagome model in the $J_2 - J_3$ plane obtained in Ref.~\cite{gong2015}.
(b-d) are static spin structure factor in momentum space for (b) the ${\bf q}=(0,0)$ phase at $J_2=0.25,J_3=0.0$, (c) the CSL at $J_2=0.25,J_3=0.25$, and (d) the KSL at $J_2=J_3=0$. The extended Brillouin zone is marked by the white dashed line.
\label{fig:Sstat}}
\end{figure*}

In this paper, we aim to understand the  DSSF based on large-scale DMRG for the KAFM and extended models with either small further-neighbor Heisenberg interactions or Dzyaloshinskii-Moriya (DM) interaction, which are both relevant to experimental material.
With these pertubative couplings, we identify characteristic features of the DSSF for different quantum phases, including a ${\bf q}=(0,0)$ magnetic order phase, a gapped chiral spin liquid (CSL), and a QSL connecting to the phase of the pure KAFM (we denote it as KSL).
In the ${\bf q} = (0,0)$ phase, the key signature of long-range magnetic order is the appearance of sharp gapless dispersive modes with the largest intensity at the corresponding magnetic wave vector.
In the CSL phase, the energy scans of the DSSF show intensity peak at finite frequency, which illustrates the emergent gapped spinon pair excitations.
In the KSL, the momentum resolved DSSF concentrates along the boundary of the extended Brillouin zone (BZ) and shows a broad maximum at the $M$ point, which are consistent with the INS results of the herbertsmithite.
In the energy scans for the KSL, the intensity of the  DSSF forms a continuum, which extends over a wide frequency range, concomitant with a pronounced intensity at low energy region.
The evidences from DSSF, including the fractionalized spinon continuum in energy scans, the sensitivity of excitation gap by imposing different boundary conditions (BCs), and by tuning DM perturbation, are in support of a QSL with gapless fractionalized spin excitations.

\begin{figure*}[!htb]
\includegraphics[width=0.275\linewidth]{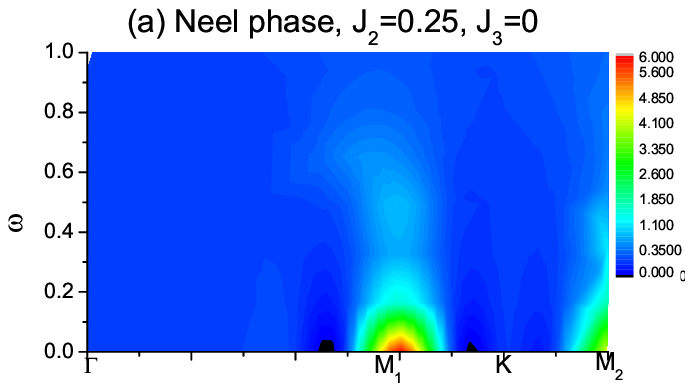}
\includegraphics[width=0.275\linewidth]{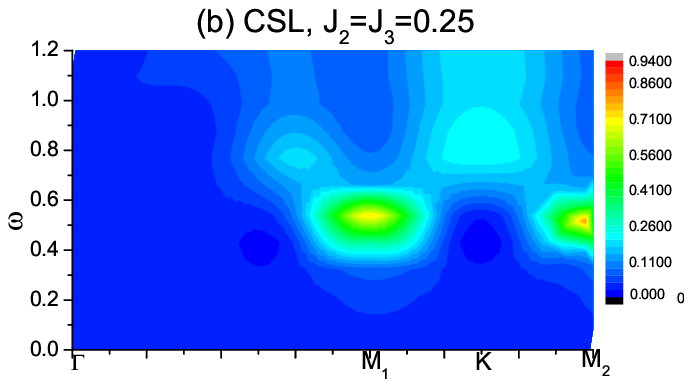}
\includegraphics[width=0.275\linewidth]{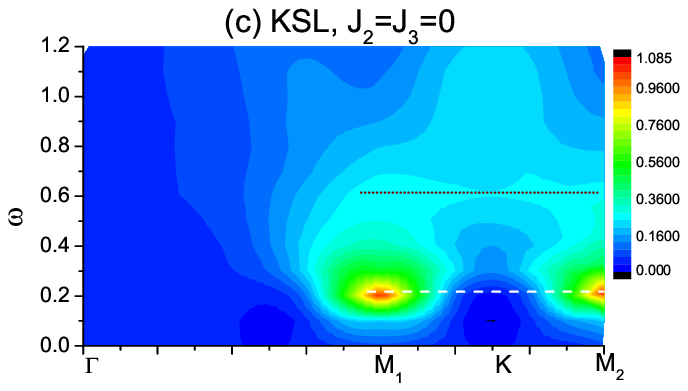}
\includegraphics[width=0.9\linewidth]{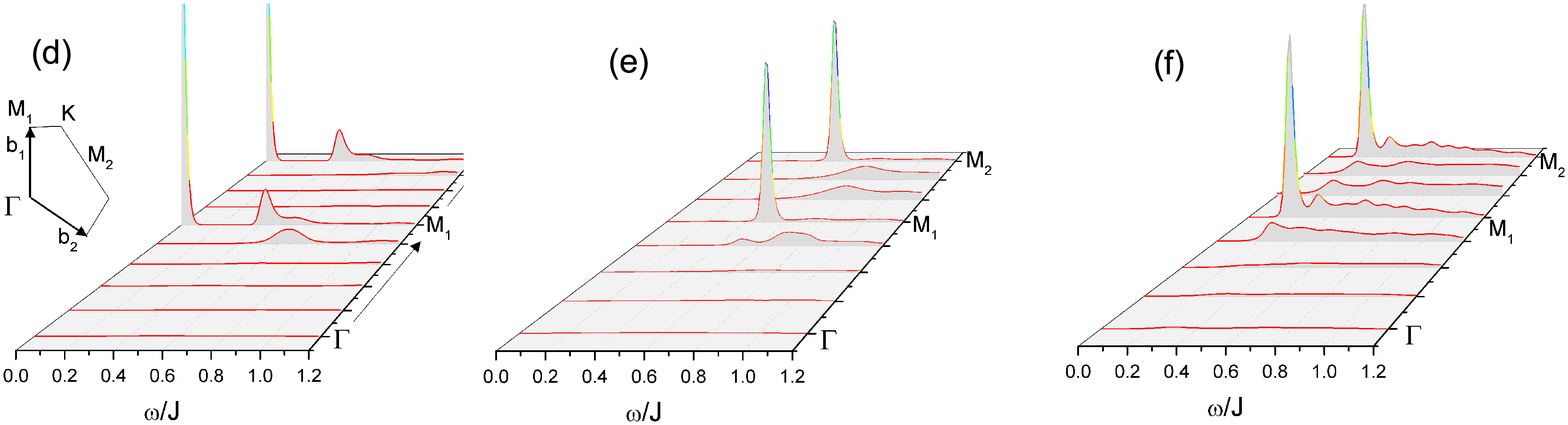}
\caption{Dynamic spin structure factor in different quantum phases.
(a-c) Contour plots of the DSSF as a function of energy and momentum for (a) $q=(0,0)$ phase at $J_2=0.25,J_3=0.0$, (b) CSL at $J_2=0.25,J_3=0.25$, and (c) KSL at $J_2=J_3=0$. The white and black dashed line in (c) shows the constant energy scan at low-frequency and high-frequency region, respectively,
which can be compared with the INS observations in herbertsmithite (see \cite{SM} for details).
(d-f) The energy scans of the DSSF with the momentum along the path $\Gamma \rightarrow M_1\rightarrow K \rightarrow M_2$ in extended BZ. The intensity scales differ among the different panels.
The inset of (d) shows reciprocal vectors of kagome lattice in the Brillouin zone with denoted high-symmetry momentum points.
\label{fig:dsq}}
\end{figure*}

\textit{Model and Method.---}
We study the spin-$1/2$ KAFM with further-neighbor antiferromagnetic interactions
\begin{equation}\label{eq:ham}
H=J_1\sum_{\langle i,j\rangle} \mathbf S_i \cdot \mathbf S_j + J_2\sum_{\langle\langle i,j\rangle\rangle} \mathbf S_i \cdot \mathbf S_j+
J_3\sum_{\langle\langle\langle i,j\rangle\rangle\rangle} \mathbf S_i \cdot \mathbf S_j,
\end{equation}
where $J_1,J_2,J_3$ are the first-, second-, and third-neighbor couplings
($J_3$ is the coupling inside the hexagon and we take $J_1 = 1$ as the energy scale).
The previously obtained DMRG phase diagram is shown in  Fig.~\ref{fig:Sstat}(a)~\cite{gong2015}.
Different neighbor phases surround the KSL sitting near the $J_1$ point, including a ${\bf q} = (0,0)$ magnetic order phase, a gapped CSL phase, and a valence-bond solid phase.
In this study, we develop a DMRG program to calculate dynamic structure factor~\cite{white1992,white1999,jeck2002}, which can apply to general strongly correlated systems.
We consider cylinder geometry with closed boundary in the $y$ direction and open boundary in the $x$ direction, with the number of sites $N = 3 \times L_x \times L_y$ $(L_x \gg L_y)$, where $L_x$ and $L_y$ are the numbers of unit cells along the $x$ and $y$ directions, respectively.
We first obtain the ground state of a long cylinder, and then target the dynamical properties by sweeping the middle $L_y \times L_y$ unit cells to avoid edge excitations (see Sec.~\ref{app:dmrg} of~\cite{SM} for details).
Most of calculations are performed on the $L_y=4$ cylinder. For the ${\bf q}=(0,0)$ phase, we obtain well converged results also for $L_y=6$.
We first present the static spin structure factor that is defined as
\begin{eqnarray*}
\mathcal{S}(\mathbf{Q})=\langle S^z(\mathbf{-Q}) S^z(\mathbf{Q}) \rangle =
\frac{1}{N} \sum_{i,j} e^{i\mathbf{Q}\cdot (\mathbf{r}_i-\mathbf{r}_j)}\langle S^z_i S^z_j \rangle,
\end{eqnarray*}
where the wave vector $\mathbf{Q}=(q_1,q_2)=q_1\vec b_1+q_2 \vec b_2$ in the BZ is defined by reciprocal vectors $\vec b_{1,2}$ (see Fig.~\ref{fig:dsq}(d)).
In Fig.~\ref{fig:Sstat}(b), $\mathcal S(\mathbf Q)$ shows sharp peaks at the $M$ points, showing a ${\bf q}=(0,0)$ magnetic order~\cite{Kolley2015}.
In the nonmagnetic phases $\mathcal S(\mathbf Q)$ is featureless as shown in Fig.~\ref{fig:Sstat}(c-d).
In the KSL phase, $\mathcal S(\mathbf Q)$ concentrates along the boundary of the extended BZ and shows broad maximum near the $M$ point, which agree with the features of the INS data of herbertsmithite~\cite{han2012}.

\begin{figure}[b]
\includegraphics[width=0.48\textwidth]{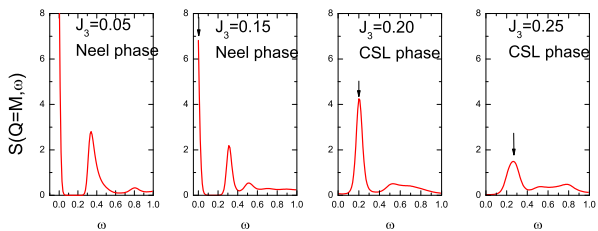}
\caption{Evolution of dynamical spin structure factor at $\mathbf Q=M$ point, by varying $J_3$.
By decreasing $J_3$ from finite to zero, It is expected that chiral spin liquid undergoes a continuous phase transition to the $q=(0,0)$ phase~\cite{gong2015}.
}\label{sfig:CSL-Neel}
\end{figure}

{\it Dynamic spin structure factor.---}
The DSSF is defined as
\begin{eqnarray*}
\mathcal{S}^{\alpha\beta}(\mathbf{Q},\omega)=-\frac{1}{\pi} \mathrm{Im}
 \langle S^\alpha(\mathbf{-Q})
\frac{1}{\omega-(H-E_{0}) +i\eta} S^\beta(\mathbf{Q}) \rangle,
\end{eqnarray*}
where $E_0$ is the ground-state energy, $\eta\rightarrow 0$ is a small smearing energy~\cite{SM}, and $\alpha, \beta$ denote spin components.
First of all, we discuss the salient features of the DSSF in different phases as shown in Fig.~\ref{fig:dsq}.
For the ${\bf q}=(0,0)$ phase (Fig.~\ref{fig:dsq}(a,d)), we observe a sharp peak at the $M$ point with $\omega=0$, serving as the key signature of the long-range magnetic order with the largest intensity at ordering wave vector.
Interestingly, we also observe a small peak with a broader and reduced weight in higher energy region, which we speculate related to two-magnon excitations.
For all other momenta, the intensity shows broad distribution in the energy scans.
In the CSL, the DSSF along the high-symmetric line is presented in Fig.~\ref{fig:dsq}(b,e), showing a fully gapped excitation branch at the $M$ point.
The extracted spin gap $0.4 J_1$ is consistent with a direct measurement of the gap in static simulation.
 For other momentum points along the boundary of the extended BZ, the DSSF has broad distribution with suppressed intensity supporting  the spin spectrum as a convolution of the
fractionalized  excitations.
Since theoretically the CSL is well described as the Laughlin state with spinons satisfying semionic statistics~\cite{kalmeyer1987}, the intensity peak at the $M$ point should  be composed of spinon pair excitations (see Ref.~\cite{SM}).
Next we turn to the KSL as shown in Fig.~\ref{fig:dsq}(c,f).
The dominant intensity of the DSSF is also carried by the momentum near the $M$ point, and the spectrum at each momentum shows broad distribution and spans a wide energy region.
For example, $\mathcal S(M,\omega)$ shows a dominant intensity at small energy and a long tail up to $\omega \approx 1.2$; the overall feature is quite different from the spectrum of the ${\bf q}=(0,0)$ phase, where the overwhelming part of the spectral weight is carried by energy $\omega=0$.
Compared with the CSL phase, here the spectrum weight moves down in energy, consistent with a reduction of spin excitation gap.
Interestingly, for the CSL phase, although the DSSF forms a continuum along the extended BZ boundary, the energy scan of dynamical spin structure factor at each momentum point shows a dominate peak structure with narrow broadening width, which supports a deconfined stable (long-lift time) spinon excitations.
Comparing with these characteristic features of the CSL, the picture shown in Fig. \ref{fig:dsq}(d,h) for the KSL indicates such a spectrum is still related to fractionalized spinon excitations with much reduced life time.
The appearance of excitation continuum in high frequency region is similar to the case of one dimensional Heisenberg model where a critical spin liquid phase has been identified as the ground state with gapless spinon excitations~\cite{mikeska2004}.
We remark that the DSSF results in the KSL phase capture the main features of the INS results of herbertsmithite, including the low-energy spectrum peak at the $M$ point and the flat spin excitations between the $M$ and $K$ points at higher energy, which would be discussed below in detail.

{\it Spinon condensation and quantum phase transition.---}
It is also interesting to study the quantum phase transition from the view of DSSF,
which reveals the dynamic driving mechanism  of  the transition. Here we study the
transition from the CSL to the ${\bf q} = (0,0)$ phase (see Ref.~\cite{SM}). From the
evolution of the DSSF at the $M$ point by tuning $J_3$, we observe the following key
features: In the CSL phase ($J_3>0.18$), with the system approaching the transition
point, the predominant peak moves towards the low-frequency regime and the peak
intensity gradually increases. After entering the N\'eel phase ($J_3<0.18$), the
predominant peak appears exactly at $\omega=0$, which is well separated from the
high-frequency excitations. The above observations indicate that
the quantum phase transition between the CSL and the  N\'eel phase
 can be understood as driven by the condense of the spinon pairs to form the
spin-1 magnon excitations.

\begin{figure}[t]
\includegraphics[width=0.5\textwidth]{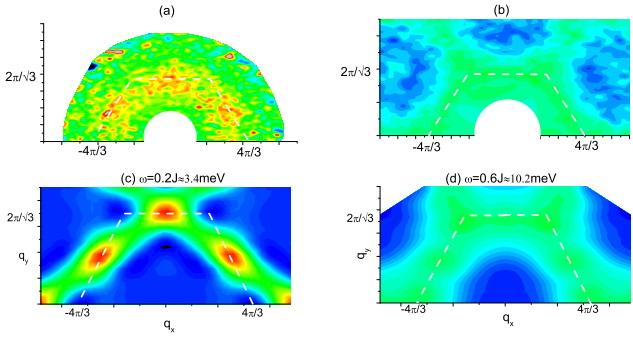}
\caption{Comparison between experimental measurements and numerical results for the DSSF.
(Top) Experimental data at fixed frequency are shown for $\omega=0.75meV$ (left) and $\omega=6meV$ (right) (The data are from experimental group \cite{han2012}).
(Bottom)Theoretical results for DSSF at fixed frequency are plotted for $\omega=0.2\approx 3meV$ (left) and $\omega=0.6\approx 10meV$ (right).
The extended Brillouin zone is indicated by the white dashed line.
}\label{contour}
\end{figure}

\begin{figure}[t]
\includegraphics[width=0.2\textwidth]{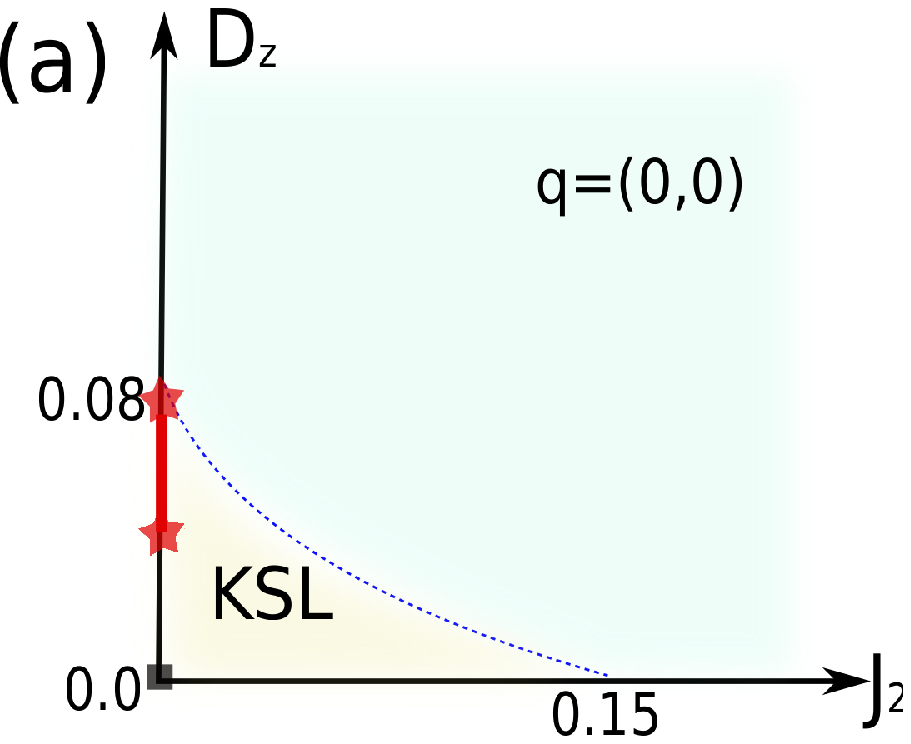}
\includegraphics[width=0.2\textwidth]{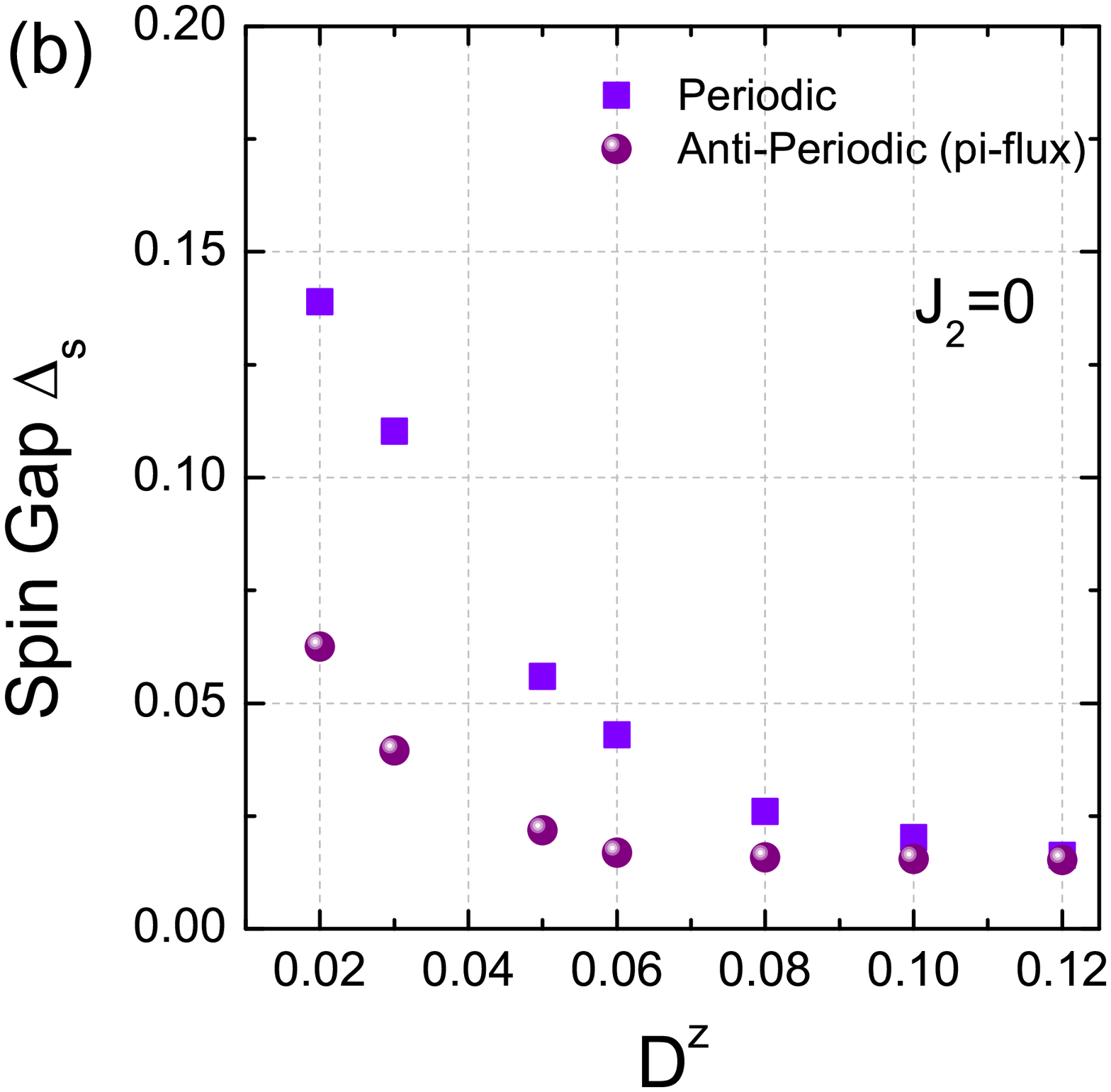}
\includegraphics[width=0.4\textwidth]{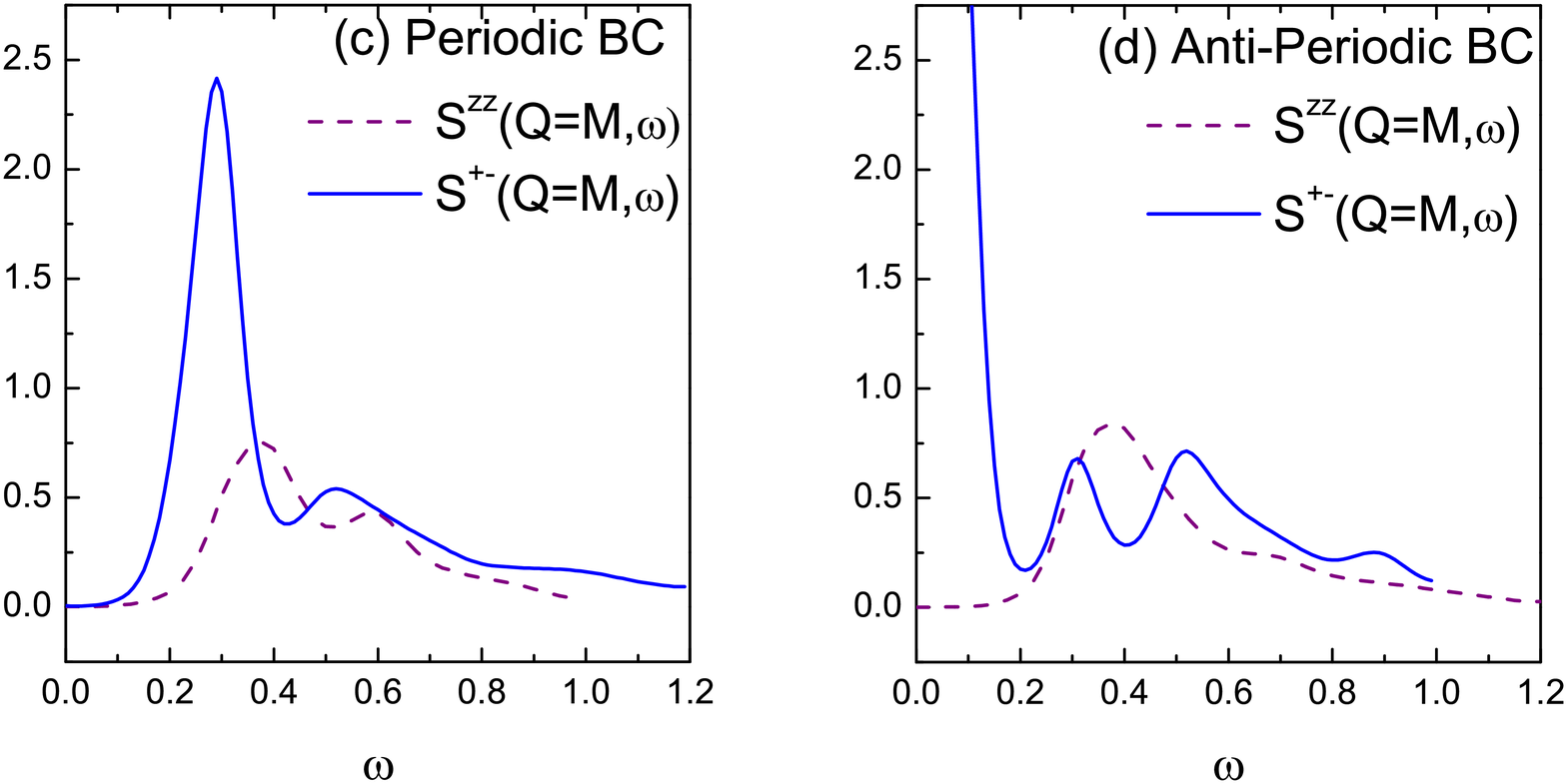}
\caption{(a) Phase diagram of KAFM by including second nearest-neighbor coupling $J_2$ and out-of-plane DM interaction $D^z$. The squared dot represents the KAFM with nearest-neighbor couplings. Red star line shows the possible parameter regime for herbertsmithite~\cite{zorko2008,han2012b}. (b) Spin gap for various $D^z$ under periodic (blue squares) and anti-periodic (purple dots) BC. The DSSF of KSL phase at $J_2=0,D^z=0.06$ under (c) periodic BC and (d) anti-periodic BC, for longitudinal mode (purple dashed line) and transverse mode (blue line).
}\label{fig:DM}
\end{figure}

{\it Connection with experiment.---}
In Fig.~\ref{contour}, we show the plots of the DSSF at constant energy, and compare our results qualitatively to the experimental data (Fig. 1 in Ref.~\cite{han2012}).
The main observation  from INS experiment is that, in the low frequency region the measured DSSF shows the peak structure around the $M$ points; while in the higher frequencies, the peak structure is smeared out, and the DSSF is almost flat distributed along the boundary of the extended BZ~\cite{han2012}.
Here we re-present two constant energy plots of the DSSF from experimental measurements in low frequency ($\omega=0.75$mev) and high frequency $\omega=6$meV, respectively as shown in Fig.~\ref{contour}(a-b).
Accordingly, we show two calculated DSSF plots at two constant energies in Fig.~\ref{contour}(c-d).
Our numerical DSSF develops peak structures around the $M$ points in low frequency and flat distribution along the boundary of the extended BZ in high frequency, respectively.
Through this comparison, we conclude that the fractionalized spinon spectrum obtained in calculations can capture the main experimental observations, both in the low frequency and the high frequency regime.
While the KAFM is generally believed as a good starting point to understand the spin-liquid-like behaviors of herbertsmithite, the spin-orbit coupling in the absence of inversion symmetry between two adjacent Cu$^{2+}$ irons yields a DM interaction $\mathbf D_{ij} \cdot (\mathbf S_i \times \mathbf S_j)$~\cite{moriya1960} in herbertsmithite.
Electron spin resonance~\cite{zorko2008} and magnetic susceptibility measurements~\cite{han2012b} suggest an out-of-plane DM interaction $D^z_{ij} \approx 0.04 \sim 0.08 J_1$.
To make a bridge between experiments and numerical simulations, we study the DSSF of the KAFM with additional DM interaction.
First of all, we show the phase diagram of the system as a function of $D^z$ and $J_2$ in Fig.~\ref{fig:DM}(a) (we set $J_3 = 0$), including the KSL and ${\bf q} = (0,0)$ phase.
We obtain the phase diagram by studying the magnetic order parameter~\cite{SM}.
In the absence of $J_2$, we find a transition at $D^z_c\approx 0.08$, slightly smaller than previous result~\cite{messio2010,cepas2008,huh2010}.
With increasing $D^z$, the spin-1 excitation gap decreases monotonically as shown in Fig.~\ref{fig:DM}(b). For $D^z < 0.08$, the spin gap depends on the BCs, similar to the DMRG results of the pure kagome model.
Since DM interaction breaks spin rotational symmetry, we calculate the DSSF in both the longitudinal and transverse modes as shown in Fig.~\ref{fig:DM}(c-d), under different BCs. It is found that the intensity distribution of the DSSF remains similar to the results of the KAFM, showing broad distribution and long tail into higher energy region.
Importantly, the low-energy excitations are governed by the transverse mode, which also shows substantial difference by tuning BC.
The dominated spectral peak in the anti-periodic BC shifts to zero frequency, showing gapless spin excitations.
These results are consistent with the KSL as a critical phase.

{\it Summary.---}
We have studied the DSSF of the spin-$1/2$ Heisenberg model on the kagome lattice with either further-neighbor or additional DM interactions using DMRG.
The DSSF of the kagome spin liquid shows different characterizations from those in the gapped CSL or in the ${\bf q} = (0,0)$ magnetic phase, which concentrates along the boundary of the extended Brillouin zone with broad maximum near the $M$ point.
In the energy scans, the dominant intensity shifts to low-energy region, and a wide spectral distribution spans to high-energy region, showing a continuum expected for a spin liquid state.
Besides, the DSSF captures the main features of the inelastic neutron scattering features of herbertsmithite.
We also propose that the DSSF could be used to characterize exotic quantum phase transitions.
{\it Note added.---} In the stage of finalizing our paper, we became aware of new preprints~\cite{Sun2018,Becker2018}, which study the dynamical properties of the $Z_2$ spin liquid in a sign free anisotropic kagome model  by the quantum Monte Carlo method.

{\it Acknowledgments.---}
W.Z. thanks for C. D. Batista, S. S. Zhang, Z. T. Wang and Y. C. He for insightful discussion.
W.Z. also thanks T. Han for discussing experimental data.
This work was supported by the U.S. Department of Energy  (DOE) through Los Alamos National Laboratory LDRD Program (W.Z.), the DOE Office of Basic Energy Sciences under the grant No. DE-FG02-06ER46305 (S.S.G., D.N.S), and the start-up funding support from Beihang University (S.S.G.).

\bibliography{dynamic}

\begin{thebibliography}{57}%
\makeatletter
\providecommand \@ifxundefined [1]{%
 \@ifx{#1\undefined}
}%
\providecommand \@ifnum [1]{%
 \ifnum #1\expandafter \@firstoftwo
 \else \expandafter \@secondoftwo
 \fi
}%
\providecommand \@ifx [1]{%
 \ifx #1\expandafter \@firstoftwo
 \else \expandafter \@secondoftwo
 \fi
}%
\providecommand \natexlab [1]{#1}%
\providecommand \enquote  [1]{``#1''}%
\providecommand \bibnamefont  [1]{#1}%
\providecommand \bibfnamefont [1]{#1}%
\providecommand \citenamefont [1]{#1}%
\providecommand \href@noop [0]{\@secondoftwo}%
\providecommand \href [0]{\begingroup \@sanitize@url \@href}%
\providecommand \@href[1]{\@@startlink{#1}\@@href}%
\providecommand \@@href[1]{\endgroup#1\@@endlink}%
\providecommand \@sanitize@url [0]{\catcode `\\12\catcode `\$12\catcode
  `\&12\catcode `\#12\catcode `\^12\catcode `\_12\catcode `\%12\relax}%
\providecommand \@@startlink[1]{}%
\providecommand \@@endlink[0]{}%
\providecommand \url  [0]{\begingroup\@sanitize@url \@url }%
\providecommand \@url [1]{\endgroup\@href {#1}{\urlprefix }}%
\providecommand \urlprefix  [0]{URL }%
\providecommand \Eprint [0]{\href }%
\providecommand \doibase [0]{http://dx.doi.org/}%
\providecommand \selectlanguage [0]{\@gobble}%
\providecommand \bibinfo  [0]{\@secondoftwo}%
\providecommand \bibfield  [0]{\@secondoftwo}%
\providecommand \translation [1]{[#1]}%
\providecommand \BibitemOpen [0]{}%
\providecommand \bibitemStop [0]{}%
\providecommand \bibitemNoStop [0]{.\EOS\space}%
\providecommand \EOS [0]{\spacefactor3000\relax}%
\providecommand \BibitemShut  [1]{\csname bibitem#1\endcsname}%
\let\auto@bib@innerbib\@empty
\bibitem [{\citenamefont {{Balents}}(2010)}]{Leon2010}%
  \BibitemOpen
  \bibfield  {author} {\bibinfo {author} {\bibfnamefont {L.}~\bibnamefont
  {{Balents}}},\ }\bibfield  {title} {\enquote {\bibinfo {title} {{Spin liquids
  in frustrated magnets}},}\ }\href {\doibase 10.1038/nature08917} {\bibfield
  {journal} {\bibinfo  {journal} {\nat}\ }\textbf {\bibinfo {volume} {464}},\
  \bibinfo {pages} {199--208} (\bibinfo {year} {2010})}\BibitemShut {NoStop}%
\bibitem [{\citenamefont {Lee}\ \emph {et~al.}(2006)\citenamefont {Lee},
  \citenamefont {Nagaosa},\ and\ \citenamefont {Wen}}]{PALee2006}%
  \BibitemOpen
  \bibfield  {author} {\bibinfo {author} {\bibfnamefont {Patrick~A.}\
  \bibnamefont {Lee}}, \bibinfo {author} {\bibfnamefont {Naoto}\ \bibnamefont
  {Nagaosa}}, \ and\ \bibinfo {author} {\bibfnamefont {Xiao-Gang}\ \bibnamefont
  {Wen}},\ }\bibfield  {title} {\enquote {\bibinfo {title} {Doping a mott
  insulator: Physics of high-temperature superconductivity},}\ }\href {\doibase
  10.1103/RevModPhys.78.17} {\bibfield  {journal} {\bibinfo  {journal} {Rev.
  Mod. Phys.}\ }\textbf {\bibinfo {volume} {78}},\ \bibinfo {pages} {17--85}
  (\bibinfo {year} {2006})}\BibitemShut {NoStop}%
\bibitem [{\citenamefont {Savary}\ and\ \citenamefont
  {Balents}(2016)}]{Leon2017}%
  \BibitemOpen
  \bibfield  {author} {\bibinfo {author} {\bibfnamefont {Lucile}\ \bibnamefont
  {Savary}}\ and\ \bibinfo {author} {\bibfnamefont {Leon}\ \bibnamefont
  {Balents}},\ }\bibfield  {title} {\enquote {\bibinfo {title} {Quantum spin
  liquids: a review},}\ }\href
  {http://iopscience.iop.org/article/10.1088/0034-4885/80/1/016502/meta}
  {\bibfield  {journal} {\bibinfo  {journal} {Reports on Progress in Physics}\
  }\textbf {\bibinfo {volume} {80}},\ \bibinfo {pages} {016502} (\bibinfo
  {year} {2016})}\BibitemShut {NoStop}%
\bibitem [{\citenamefont {Wen}(1989)}]{Wen1989}%
  \BibitemOpen
  \bibfield  {author} {\bibinfo {author} {\bibfnamefont {X.~G.}\ \bibnamefont
  {Wen}},\ }\bibfield  {title} {\enquote {\bibinfo {title} {Vacuum degeneracy
  of chiral spin states in compactified space},}\ }\href {\doibase
  10.1103/PhysRevB.40.7387} {\bibfield  {journal} {\bibinfo  {journal} {Phys.
  Rev. B}\ }\textbf {\bibinfo {volume} {40}},\ \bibinfo {pages} {7387--7390}
  (\bibinfo {year} {1989})}\BibitemShut {NoStop}%
\bibitem [{\citenamefont {Wen}\ and\ \citenamefont {Niu}(1990)}]{WenNiu1990}%
  \BibitemOpen
  \bibfield  {author} {\bibinfo {author} {\bibfnamefont {X.~G.}\ \bibnamefont
  {Wen}}\ and\ \bibinfo {author} {\bibfnamefont {Q.}~\bibnamefont {Niu}},\
  }\bibfield  {title} {\enquote {\bibinfo {title} {Ground-state degeneracy of
  the fractional quantum hall states in the presence of a random potential and
  on high-genus riemann surfaces},}\ }\href {\doibase 10.1103/PhysRevB.41.9377}
  {\bibfield  {journal} {\bibinfo  {journal} {Phys. Rev. B}\ }\textbf {\bibinfo
  {volume} {41}},\ \bibinfo {pages} {9377--9396} (\bibinfo {year}
  {1990})}\BibitemShut {NoStop}%
\bibitem [{\citenamefont {Wen}(1991)}]{Wen1991}%
  \BibitemOpen
  \bibfield  {author} {\bibinfo {author} {\bibfnamefont {X.~G.}\ \bibnamefont
  {Wen}},\ }\bibfield  {title} {\enquote {\bibinfo {title} {Mean-field theory
  of spin-liquid states with finite energy gap and topological orders},}\
  }\href {\doibase 10.1103/PhysRevB.44.2664} {\bibfield  {journal} {\bibinfo
  {journal} {Phys. Rev. B}\ }\textbf {\bibinfo {volume} {44}},\ \bibinfo
  {pages} {2664--2672} (\bibinfo {year} {1991})}\BibitemShut {NoStop}%
\bibitem [{\citenamefont {Read}\ and\ \citenamefont
  {Sachdev}(1991)}]{Read1991}%
  \BibitemOpen
  \bibfield  {author} {\bibinfo {author} {\bibfnamefont {N.}~\bibnamefont
  {Read}}\ and\ \bibinfo {author} {\bibfnamefont {Subir}\ \bibnamefont
  {Sachdev}},\ }\bibfield  {title} {\enquote {\bibinfo {title} {Large-n
  expansion for frustrated quantum antiferromagnets},}\ }\href {\doibase
  10.1103/PhysRevLett.66.1773} {\bibfield  {journal} {\bibinfo  {journal}
  {Phys. Rev. Lett.}\ }\textbf {\bibinfo {volume} {66}},\ \bibinfo {pages}
  {1773--1776} (\bibinfo {year} {1991})}\BibitemShut {NoStop}%
\bibitem [{\citenamefont {Mendels}\ \emph {et~al.}(2007)\citenamefont
  {Mendels}, \citenamefont {Bert}, \citenamefont {de~Vries}, \citenamefont
  {Olariu}, \citenamefont {Harrison}, \citenamefont {Duc}, \citenamefont
  {Trombe}, \citenamefont {Lord}, \citenamefont {Amato},\ and\ \citenamefont
  {Baines}}]{mendels2007}%
  \BibitemOpen
  \bibfield  {author} {\bibinfo {author} {\bibfnamefont {P.}~\bibnamefont
  {Mendels}}, \bibinfo {author} {\bibfnamefont {F.}~\bibnamefont {Bert}},
  \bibinfo {author} {\bibfnamefont {M.~A.}\ \bibnamefont {de~Vries}}, \bibinfo
  {author} {\bibfnamefont {A.}~\bibnamefont {Olariu}}, \bibinfo {author}
  {\bibfnamefont {A.}~\bibnamefont {Harrison}}, \bibinfo {author}
  {\bibfnamefont {F.}~\bibnamefont {Duc}}, \bibinfo {author} {\bibfnamefont
  {J.~C.}\ \bibnamefont {Trombe}}, \bibinfo {author} {\bibfnamefont {J.~S.}\
  \bibnamefont {Lord}}, \bibinfo {author} {\bibfnamefont {A.}~\bibnamefont
  {Amato}}, \ and\ \bibinfo {author} {\bibfnamefont {C.}~\bibnamefont
  {Baines}},\ }\bibfield  {title} {\enquote {\bibinfo {title} {Quantum
  magnetism in the paratacamite family: Towards an ideal kagom\'e lattice},}\
  }\href {\doibase 10.1103/PhysRevLett.98.077204} {\bibfield  {journal}
  {\bibinfo  {journal} {Phys. Rev. Lett.}\ }\textbf {\bibinfo {volume} {98}},\
  \bibinfo {pages} {077204} (\bibinfo {year} {2007})}\BibitemShut {NoStop}%
\bibitem [{\citenamefont {Helton}\ \emph {et~al.}(2007)\citenamefont {Helton},
  \citenamefont {Matan}, \citenamefont {Shores}, \citenamefont {Nytko},
  \citenamefont {Bartlett}, \citenamefont {Yoshida}, \citenamefont {Takano},
  \citenamefont {Suslov}, \citenamefont {Qiu}, \citenamefont {Chung},
  \citenamefont {Nocera},\ and\ \citenamefont {Lee}}]{helton2007}%
  \BibitemOpen
  \bibfield  {author} {\bibinfo {author} {\bibfnamefont {J.~S.}\ \bibnamefont
  {Helton}}, \bibinfo {author} {\bibfnamefont {K.}~\bibnamefont {Matan}},
  \bibinfo {author} {\bibfnamefont {M.~P.}\ \bibnamefont {Shores}}, \bibinfo
  {author} {\bibfnamefont {E.~A.}\ \bibnamefont {Nytko}}, \bibinfo {author}
  {\bibfnamefont {B.~M.}\ \bibnamefont {Bartlett}}, \bibinfo {author}
  {\bibfnamefont {Y.}~\bibnamefont {Yoshida}}, \bibinfo {author} {\bibfnamefont
  {Y.}~\bibnamefont {Takano}}, \bibinfo {author} {\bibfnamefont
  {A.}~\bibnamefont {Suslov}}, \bibinfo {author} {\bibfnamefont
  {Y.}~\bibnamefont {Qiu}}, \bibinfo {author} {\bibfnamefont {J.-H.}\
  \bibnamefont {Chung}}, \bibinfo {author} {\bibfnamefont {D.~G.}\ \bibnamefont
  {Nocera}}, \ and\ \bibinfo {author} {\bibfnamefont {Y.~S.}\ \bibnamefont
  {Lee}},\ }\bibfield  {title} {\enquote {\bibinfo {title} {Spin dynamics of
  the spin-$1/2$ kagome lattice antiferromagnet
  ${\mathrm{zncu}}_{3}(\mathrm{OH}{)}_{6}{\mathrm{cl}}_{2}$},}\ }\href
  {\doibase 10.1103/PhysRevLett.98.107204} {\bibfield  {journal} {\bibinfo
  {journal} {Phys. Rev. Lett.}\ }\textbf {\bibinfo {volume} {98}},\ \bibinfo
  {pages} {107204} (\bibinfo {year} {2007})}\BibitemShut {NoStop}%
\bibitem [{\citenamefont {{Han}}\ \emph {et~al.}(2012)\citenamefont {{Han}},
  \citenamefont {{Helton}}, \citenamefont {{Chu}}, \citenamefont {{Nocera}},
  \citenamefont {{Rodriguez-Rivera}}, \citenamefont {{Broholm}},\ and\
  \citenamefont {{Lee}}}]{han2012}%
  \BibitemOpen
  \bibfield  {author} {\bibinfo {author} {\bibfnamefont {T.-H.}\ \bibnamefont
  {{Han}}}, \bibinfo {author} {\bibfnamefont {J.~S.}\ \bibnamefont {{Helton}}},
  \bibinfo {author} {\bibfnamefont {S.}~\bibnamefont {{Chu}}}, \bibinfo
  {author} {\bibfnamefont {D.~G.}\ \bibnamefont {{Nocera}}}, \bibinfo {author}
  {\bibfnamefont {J.~A.}\ \bibnamefont {{Rodriguez-Rivera}}}, \bibinfo {author}
  {\bibfnamefont {C.}~\bibnamefont {{Broholm}}}, \ and\ \bibinfo {author}
  {\bibfnamefont {Y.~S.}\ \bibnamefont {{Lee}}},\ }\bibfield  {title} {\enquote
  {\bibinfo {title} {{Fractionalized excitations in the spin-liquid state of a
  kagome-lattice antiferromagnet}},}\ }\href {\doibase 10.1038/nature11659}
  {\bibfield  {journal} {\bibinfo  {journal} {\nat}\ }\textbf {\bibinfo
  {volume} {492}},\ \bibinfo {pages} {406--410} (\bibinfo {year}
  {2012})}\BibitemShut {NoStop}%
\bibitem [{\citenamefont {{Fu}}\ \emph {et~al.}(2015)\citenamefont {{Fu}},
  \citenamefont {{Imai}}, \citenamefont {{Han}},\ and\ \citenamefont
  {{Lee}}}]{fu2015}%
  \BibitemOpen
  \bibfield  {author} {\bibinfo {author} {\bibfnamefont {M.}~\bibnamefont
  {{Fu}}}, \bibinfo {author} {\bibfnamefont {T.}~\bibnamefont {{Imai}}},
  \bibinfo {author} {\bibfnamefont {T.-H.}\ \bibnamefont {{Han}}}, \ and\
  \bibinfo {author} {\bibfnamefont {Y.~S.}\ \bibnamefont {{Lee}}},\ }\bibfield
  {title} {\enquote {\bibinfo {title} {{Evidence for a gapped spin-liquid
  ground state in a kagome Heisenberg antiferromagnet}},}\ }\href {\doibase
  10.1126/science.aab2120} {\bibfield  {journal} {\bibinfo  {journal}
  {Science}\ }\textbf {\bibinfo {volume} {350}},\ \bibinfo {pages} {655--658}
  (\bibinfo {year} {2015})}\BibitemShut {NoStop}%
\bibitem [{\citenamefont {Yamashita}\ \emph {et~al.}(2008)\citenamefont
  {Yamashita}, \citenamefont {Nakazawa}, \citenamefont {Oguni}, \citenamefont
  {Oshima}, \citenamefont {Nojiri}, \citenamefont {Shimizu}, \citenamefont
  {Miyagawa},\ and\ \citenamefont {Kanoda}}]{yamashita2008}%
  \BibitemOpen
  \bibfield  {author} {\bibinfo {author} {\bibfnamefont {Satoshi}\ \bibnamefont
  {Yamashita}}, \bibinfo {author} {\bibfnamefont {Yasuhiro}\ \bibnamefont
  {Nakazawa}}, \bibinfo {author} {\bibfnamefont {Masaharu}\ \bibnamefont
  {Oguni}}, \bibinfo {author} {\bibfnamefont {Yugo}\ \bibnamefont {Oshima}},
  \bibinfo {author} {\bibfnamefont {Hiroyuki}\ \bibnamefont {Nojiri}}, \bibinfo
  {author} {\bibfnamefont {Yasuhiro}\ \bibnamefont {Shimizu}}, \bibinfo
  {author} {\bibfnamefont {Kazuya}\ \bibnamefont {Miyagawa}}, \ and\ \bibinfo
  {author} {\bibfnamefont {Kazushi}\ \bibnamefont {Kanoda}},\ }\bibfield
  {title} {\enquote {\bibinfo {title} {Thermodynamic properties of a spin-1/2
  spin-liquid state in a $\kappa$-type organic salt},}\ }\href
  {http://www.nature.com/nphys/journal/v4/n6/abs/nphys942.html} {\bibfield
  {journal} {\bibinfo  {journal} {Nature Physics}\ }\textbf {\bibinfo {volume}
  {4}},\ \bibinfo {pages} {459--462} (\bibinfo {year} {2008})}\BibitemShut
  {NoStop}%
\bibitem [{\citenamefont {Shimizu}\ \emph {et~al.}(2003)\citenamefont
  {Shimizu}, \citenamefont {Miyagawa}, \citenamefont {Kanoda}, \citenamefont
  {Maesato},\ and\ \citenamefont {Saito}}]{shimizu2003}%
  \BibitemOpen
  \bibfield  {author} {\bibinfo {author} {\bibfnamefont {Y.}~\bibnamefont
  {Shimizu}}, \bibinfo {author} {\bibfnamefont {K.}~\bibnamefont {Miyagawa}},
  \bibinfo {author} {\bibfnamefont {K.}~\bibnamefont {Kanoda}}, \bibinfo
  {author} {\bibfnamefont {M.}~\bibnamefont {Maesato}}, \ and\ \bibinfo
  {author} {\bibfnamefont {G.}~\bibnamefont {Saito}},\ }\bibfield  {title}
  {\enquote {\bibinfo {title} {Spin liquid state in an organic mott insulator
  with a triangular lattice},}\ }\href {\doibase 10.1103/PhysRevLett.91.107001}
  {\bibfield  {journal} {\bibinfo  {journal} {Phys. Rev. Lett.}\ }\textbf
  {\bibinfo {volume} {91}},\ \bibinfo {pages} {107001} (\bibinfo {year}
  {2003})}\BibitemShut {NoStop}%
\bibitem [{\citenamefont {Kurosaki}\ \emph {et~al.}(2005)\citenamefont
  {Kurosaki}, \citenamefont {Shimizu}, \citenamefont {Miyagawa}, \citenamefont
  {Kanoda},\ and\ \citenamefont {Saito}}]{kurosaki2005}%
  \BibitemOpen
  \bibfield  {author} {\bibinfo {author} {\bibfnamefont {Y.}~\bibnamefont
  {Kurosaki}}, \bibinfo {author} {\bibfnamefont {Y.}~\bibnamefont {Shimizu}},
  \bibinfo {author} {\bibfnamefont {K.}~\bibnamefont {Miyagawa}}, \bibinfo
  {author} {\bibfnamefont {K.}~\bibnamefont {Kanoda}}, \ and\ \bibinfo {author}
  {\bibfnamefont {G.}~\bibnamefont {Saito}},\ }\bibfield  {title} {\enquote
  {\bibinfo {title} {Mott transition from a spin liquid to a fermi liquid in
  the spin-frustrated organic conductor $\kappa$-(et)$_2$cu$_2$(cn)$_3$},}\
  }\href {\doibase 10.1103/PhysRevLett.95.177001} {\bibfield  {journal}
  {\bibinfo  {journal} {Phys. Rev. Lett.}\ }\textbf {\bibinfo {volume} {95}},\
  \bibinfo {pages} {177001} (\bibinfo {year} {2005})}\BibitemShut {NoStop}%
\bibitem [{\citenamefont {Norman}(2016)}]{norman2016}%
  \BibitemOpen
  \bibfield  {author} {\bibinfo {author} {\bibfnamefont {M.~R.}\ \bibnamefont
  {Norman}},\ }\bibfield  {title} {\enquote {\bibinfo {title}
  {Colloquium:herbertsmithite and the search for the quantum spin liquid},}\
  }\href {\doibase 10.1103/RevModPhys.88.041002} {\bibfield  {journal}
  {\bibinfo  {journal} {Rev. Mod. Phys.}\ }\textbf {\bibinfo {volume} {88}},\
  \bibinfo {pages} {041002} (\bibinfo {year} {2016})}\BibitemShut {NoStop}%
\bibitem [{\citenamefont {Shimokawa}\ \emph {et~al.}(2015)\citenamefont
  {Shimokawa}, \citenamefont {Watanabe},\ and\ \citenamefont
  {Kawamura}}]{shimokawa2015}%
  \BibitemOpen
  \bibfield  {author} {\bibinfo {author} {\bibfnamefont {Tokuro}\ \bibnamefont
  {Shimokawa}}, \bibinfo {author} {\bibfnamefont {Ken}\ \bibnamefont
  {Watanabe}}, \ and\ \bibinfo {author} {\bibfnamefont {Hikaru}\ \bibnamefont
  {Kawamura}},\ }\bibfield  {title} {\enquote {\bibinfo {title} {Static and
  dynamical spin correlations of the $s=\frac{1}{2}$ random-bond
  antiferromagnetic heisenberg model on the triangular and kagome lattices},}\
  }\href {\doibase 10.1103/PhysRevB.92.134407} {\bibfield  {journal} {\bibinfo
  {journal} {Phys. Rev. B}\ }\textbf {\bibinfo {volume} {92}},\ \bibinfo
  {pages} {134407} (\bibinfo {year} {2015})}\BibitemShut {NoStop}%
\bibitem [{\citenamefont {Sachdev}(1992)}]{sachdev1992}%
  \BibitemOpen
  \bibfield  {author} {\bibinfo {author} {\bibfnamefont {Subir}\ \bibnamefont
  {Sachdev}},\ }\bibfield  {title} {\enquote {\bibinfo {title} {Kagome and
  triangular lattice heisenberg antiferromagnets: Ordering from quantum
  fluctuations and quantum-disordered ground states with unconfined bosonic
  spinons},}\ }\href {\doibase 10.1103/PhysRevB.45.12377} {\bibfield  {journal}
  {\bibinfo  {journal} {Phys. Rev. B}\ }\textbf {\bibinfo {volume} {45}},\
  \bibinfo {pages} {12377--12396} (\bibinfo {year} {1992})}\BibitemShut
  {NoStop}%
\bibitem [{\citenamefont {Waldtmann}\ \emph {et~al.}(1998)\citenamefont
  {Waldtmann}, \citenamefont {Everts}, \citenamefont {Bernu}, \citenamefont
  {Lhuillier}, \citenamefont {Sindzingre}, \citenamefont {Lecheminant},\ and\
  \citenamefont {Pierre}}]{waldtmann1998}%
  \BibitemOpen
  \bibfield  {author} {\bibinfo {author} {\bibfnamefont {C}~\bibnamefont
  {Waldtmann}}, \bibinfo {author} {\bibfnamefont {H-U}\ \bibnamefont {Everts}},
  \bibinfo {author} {\bibfnamefont {B}~\bibnamefont {Bernu}}, \bibinfo {author}
  {\bibfnamefont {C}~\bibnamefont {Lhuillier}}, \bibinfo {author}
  {\bibfnamefont {P}~\bibnamefont {Sindzingre}}, \bibinfo {author}
  {\bibfnamefont {P}~\bibnamefont {Lecheminant}}, \ and\ \bibinfo {author}
  {\bibfnamefont {L}~\bibnamefont {Pierre}},\ }\bibfield  {title} {\enquote
  {\bibinfo {title} {First excitations of the spin 1/2 heisenberg
  antiferromagnet on the kagom{\'e} lattice},}\ }\href
  {https://link.springer.com/article/10.1007%2Fs100510050274?LI=true}
  {\bibfield  {journal} {\bibinfo  {journal} {The European Physical Journal
  B-Condensed Matter and Complex Systems}\ }\textbf {\bibinfo {volume} {2}},\
  \bibinfo {pages} {501--507} (\bibinfo {year} {1998})}\BibitemShut {NoStop}%
\bibitem [{\citenamefont {Ran}\ \emph {et~al.}(2007)\citenamefont {Ran},
  \citenamefont {Hermele}, \citenamefont {Lee},\ and\ \citenamefont
  {Wen}}]{Ran2007}%
  \BibitemOpen
  \bibfield  {author} {\bibinfo {author} {\bibfnamefont {Ying}\ \bibnamefont
  {Ran}}, \bibinfo {author} {\bibfnamefont {Michael}\ \bibnamefont {Hermele}},
  \bibinfo {author} {\bibfnamefont {Patrick~A.}\ \bibnamefont {Lee}}, \ and\
  \bibinfo {author} {\bibfnamefont {Xiao-Gang}\ \bibnamefont {Wen}},\
  }\bibfield  {title} {\enquote {\bibinfo {title} {Projected-wave-function
  study of the spin-$1/2$ heisenberg model on the kagom\'e lattice},}\ }\href
  {\doibase 10.1103/PhysRevLett.98.117205} {\bibfield  {journal} {\bibinfo
  {journal} {Phys. Rev. Lett.}\ }\textbf {\bibinfo {volume} {98}},\ \bibinfo
  {pages} {117205} (\bibinfo {year} {2007})}\BibitemShut {NoStop}%
\bibitem [{\citenamefont {Hermele}\ \emph {et~al.}(2008)\citenamefont
  {Hermele}, \citenamefont {Ran}, \citenamefont {Lee},\ and\ \citenamefont
  {Wen}}]{Hermele2008}%
  \BibitemOpen
  \bibfield  {author} {\bibinfo {author} {\bibfnamefont {Michael}\ \bibnamefont
  {Hermele}}, \bibinfo {author} {\bibfnamefont {Ying}\ \bibnamefont {Ran}},
  \bibinfo {author} {\bibfnamefont {Patrick~A.}\ \bibnamefont {Lee}}, \ and\
  \bibinfo {author} {\bibfnamefont {Xiao-Gang}\ \bibnamefont {Wen}},\
  }\bibfield  {title} {\enquote {\bibinfo {title} {Properties of an algebraic
  spin liquid on the kagome lattice},}\ }\href {\doibase
  10.1103/PhysRevB.77.224413} {\bibfield  {journal} {\bibinfo  {journal} {Phys.
  Rev. B}\ }\textbf {\bibinfo {volume} {77}},\ \bibinfo {pages} {224413}
  (\bibinfo {year} {2008})}\BibitemShut {NoStop}%
\bibitem [{\citenamefont {Iqbal}\ \emph {et~al.}(2011)\citenamefont {Iqbal},
  \citenamefont {Becca},\ and\ \citenamefont {Poilblanc}}]{Iqbal2011}%
  \BibitemOpen
  \bibfield  {author} {\bibinfo {author} {\bibfnamefont {Yasir}\ \bibnamefont
  {Iqbal}}, \bibinfo {author} {\bibfnamefont {Federico}\ \bibnamefont {Becca}},
  \ and\ \bibinfo {author} {\bibfnamefont {Didier}\ \bibnamefont {Poilblanc}},\
  }\bibfield  {title} {\enquote {\bibinfo {title} {Projected wave function
  study of ${\mathbb{z}}_{2}$ spin liquids on the kagome lattice for the
  spin-$\frac{1}{2}$ quantum heisenberg antiferromagnet},}\ }\href {\doibase
  10.1103/PhysRevB.84.020407} {\bibfield  {journal} {\bibinfo  {journal} {Phys.
  Rev. B}\ }\textbf {\bibinfo {volume} {84}},\ \bibinfo {pages} {020407}
  (\bibinfo {year} {2011})}\BibitemShut {NoStop}%
\bibitem [{\citenamefont {Iqbal}\ \emph {et~al.}(2013)\citenamefont {Iqbal},
  \citenamefont {Becca}, \citenamefont {Sorella},\ and\ \citenamefont
  {Poilblanc}}]{Iqbal2013}%
  \BibitemOpen
  \bibfield  {author} {\bibinfo {author} {\bibfnamefont {Yasir}\ \bibnamefont
  {Iqbal}}, \bibinfo {author} {\bibfnamefont {Federico}\ \bibnamefont {Becca}},
  \bibinfo {author} {\bibfnamefont {Sandro}\ \bibnamefont {Sorella}}, \ and\
  \bibinfo {author} {\bibfnamefont {Didier}\ \bibnamefont {Poilblanc}},\
  }\bibfield  {title} {\enquote {\bibinfo {title} {Gapless spin-liquid phase in
  the kagome spin-$\frac{1}{2}$ heisenberg antiferromagnet},}\ }\href {\doibase
  10.1103/PhysRevB.87.060405} {\bibfield  {journal} {\bibinfo  {journal} {Phys.
  Rev. B}\ }\textbf {\bibinfo {volume} {87}},\ \bibinfo {pages} {060405}
  (\bibinfo {year} {2013})}\BibitemShut {NoStop}%
\bibitem [{\citenamefont {Iqbal}\ \emph {et~al.}(2014)\citenamefont {Iqbal},
  \citenamefont {Poilblanc},\ and\ \citenamefont {Becca}}]{Iqbal2014}%
  \BibitemOpen
  \bibfield  {author} {\bibinfo {author} {\bibfnamefont {Yasir}\ \bibnamefont
  {Iqbal}}, \bibinfo {author} {\bibfnamefont {Didier}\ \bibnamefont
  {Poilblanc}}, \ and\ \bibinfo {author} {\bibfnamefont {Federico}\
  \bibnamefont {Becca}},\ }\bibfield  {title} {\enquote {\bibinfo {title}
  {Vanishing spin gap in a competing spin-liquid phase in the kagome heisenberg
  antiferromagnet},}\ }\href {\doibase 10.1103/PhysRevB.89.020407} {\bibfield
  {journal} {\bibinfo  {journal} {Phys. Rev. B}\ }\textbf {\bibinfo {volume}
  {89}},\ \bibinfo {pages} {020407} (\bibinfo {year} {2014})}\BibitemShut
  {NoStop}%
\bibitem [{\citenamefont {{Yan}}\ \emph {et~al.}(2011)\citenamefont {{Yan}},
  \citenamefont {{Huse}},\ and\ \citenamefont {{White}}}]{Yan2011}%
  \BibitemOpen
  \bibfield  {author} {\bibinfo {author} {\bibfnamefont {S.}~\bibnamefont
  {{Yan}}}, \bibinfo {author} {\bibfnamefont {D.~A.}\ \bibnamefont {{Huse}}}, \
  and\ \bibinfo {author} {\bibfnamefont {S.~R.}\ \bibnamefont {{White}}},\
  }\bibfield  {title} {\enquote {\bibinfo {title} {{Spin-Liquid Ground State of
  the S = 1/2 Kagome Heisenberg Antiferromagnet}},}\ }\href {\doibase
  10.1126/science.1201080} {\bibfield  {journal} {\bibinfo  {journal}
  {Science}\ }\textbf {\bibinfo {volume} {332}},\ \bibinfo {pages} {1173}
  (\bibinfo {year} {2011})}\BibitemShut {NoStop}%
\bibitem [{\citenamefont {Depenbrock}\ \emph {et~al.}(2012)\citenamefont
  {Depenbrock}, \citenamefont {McCulloch},\ and\ \citenamefont
  {Schollw\"ock}}]{Depenbrock2012}%
  \BibitemOpen
  \bibfield  {author} {\bibinfo {author} {\bibfnamefont {Stefan}\ \bibnamefont
  {Depenbrock}}, \bibinfo {author} {\bibfnamefont {Ian~P.}\ \bibnamefont
  {McCulloch}}, \ and\ \bibinfo {author} {\bibfnamefont {Ulrich}\ \bibnamefont
  {Schollw\"ock}},\ }\bibfield  {title} {\enquote {\bibinfo {title} {Nature of
  the spin-liquid ground state of the $s=1/2$ heisenberg model on the kagome
  lattice},}\ }\href {\doibase 10.1103/PhysRevLett.109.067201} {\bibfield
  {journal} {\bibinfo  {journal} {Phys. Rev. Lett.}\ }\textbf {\bibinfo
  {volume} {109}},\ \bibinfo {pages} {067201} (\bibinfo {year}
  {2012})}\BibitemShut {NoStop}%
\bibitem [{\citenamefont {Jiang}\ \emph {et~al.}(2012)\citenamefont {Jiang},
  \citenamefont {Wang},\ and\ \citenamefont {Balents}}]{Jiang2012nature}%
  \BibitemOpen
  \bibfield  {author} {\bibinfo {author} {\bibfnamefont {Hong-Chen}\
  \bibnamefont {Jiang}}, \bibinfo {author} {\bibfnamefont {Zhenghan}\
  \bibnamefont {Wang}}, \ and\ \bibinfo {author} {\bibfnamefont {Leon}\
  \bibnamefont {Balents}},\ }\bibfield  {title} {\enquote {\bibinfo {title}
  {Identifying topological order by entanglement entropy},}\ }\href
  {http://www.nature.com/nphys/journal/v8/n12/full/nphys2465.html} {\bibfield
  {journal} {\bibinfo  {journal} {Nature Physics}\ }\textbf {\bibinfo {volume}
  {8}},\ \bibinfo {pages} {902--905} (\bibinfo {year} {2012})}\BibitemShut
  {NoStop}%
\bibitem [{\citenamefont {Jiang}\ \emph {et~al.}(2008)\citenamefont {Jiang},
  \citenamefont {Weng},\ and\ \citenamefont {Sheng}}]{Jiang2008}%
  \BibitemOpen
  \bibfield  {author} {\bibinfo {author} {\bibfnamefont {H.~C.}\ \bibnamefont
  {Jiang}}, \bibinfo {author} {\bibfnamefont {Z.~Y.}\ \bibnamefont {Weng}}, \
  and\ \bibinfo {author} {\bibfnamefont {D.~N.}\ \bibnamefont {Sheng}},\
  }\bibfield  {title} {\enquote {\bibinfo {title} {Density matrix
  renormalization group numerical study of the kagome antiferromagnet},}\
  }\href {\doibase 10.1103/PhysRevLett.101.117203} {\bibfield  {journal}
  {\bibinfo  {journal} {Phys. Rev. Lett.}\ }\textbf {\bibinfo {volume} {101}},\
  \bibinfo {pages} {117203} (\bibinfo {year} {2008})}\BibitemShut {NoStop}%
\bibitem [{\citenamefont {Messio}\ \emph {et~al.}(2012)\citenamefont {Messio},
  \citenamefont {Bernu},\ and\ \citenamefont {Lhuillier}}]{messio2012}%
  \BibitemOpen
  \bibfield  {author} {\bibinfo {author} {\bibfnamefont {Laura}\ \bibnamefont
  {Messio}}, \bibinfo {author} {\bibfnamefont {Bernard}\ \bibnamefont {Bernu}},
  \ and\ \bibinfo {author} {\bibfnamefont {Claire}\ \bibnamefont {Lhuillier}},\
  }\bibfield  {title} {\enquote {\bibinfo {title} {Kagome antiferromagnet: A
  chiral topological spin liquid?}}\ }\href {\doibase
  10.1103/PhysRevLett.108.207204} {\bibfield  {journal} {\bibinfo  {journal}
  {Phys. Rev. Lett.}\ }\textbf {\bibinfo {volume} {108}},\ \bibinfo {pages}
  {207204} (\bibinfo {year} {2012})}\BibitemShut {NoStop}%
\bibitem [{\citenamefont {{L{\"a}uchli}}\ \emph {et~al.}(2016)\citenamefont
  {{L{\"a}uchli}}, \citenamefont {{Sudan}},\ and\ \citenamefont
  {{Moessner}}}]{lauchli2016}%
  \BibitemOpen
  \bibfield  {author} {\bibinfo {author} {\bibfnamefont {A.~M.}\ \bibnamefont
  {{L{\"a}uchli}}}, \bibinfo {author} {\bibfnamefont {J.}~\bibnamefont
  {{Sudan}}}, \ and\ \bibinfo {author} {\bibfnamefont {R.}~\bibnamefont
  {{Moessner}}},\ }\bibfield  {title} {\enquote {\bibinfo {title} {{The $S=1/2$
  Kagome Heisenberg Antiferromagnet Revisited}},}\ }\href@noop {} {\bibfield
  {journal} {\bibinfo  {journal} {ArXiv e-prints}\ } (\bibinfo {year}
  {2016})},\ \Eprint {http://arxiv.org/abs/1611.06990} {arXiv:1611.06990
  [cond-mat.str-el]} \BibitemShut {NoStop}%
\bibitem [{\citenamefont {Mei}\ \emph {et~al.}(2017)\citenamefont {Mei},
  \citenamefont {Chen}, \citenamefont {He},\ and\ \citenamefont
  {Wen}}]{mei2016}%
  \BibitemOpen
  \bibfield  {author} {\bibinfo {author} {\bibfnamefont {Jia-Wei}\ \bibnamefont
  {Mei}}, \bibinfo {author} {\bibfnamefont {Ji-Yao}\ \bibnamefont {Chen}},
  \bibinfo {author} {\bibfnamefont {Huan}\ \bibnamefont {He}}, \ and\ \bibinfo
  {author} {\bibfnamefont {Xiao-Gang}\ \bibnamefont {Wen}},\ }\bibfield
  {title} {\enquote {\bibinfo {title} {Gapped spin liquid with
  ${\mathbb{z}}_{2}$ topological order for the kagome heisenberg model},}\
  }\href {\doibase 10.1103/PhysRevB.95.235107} {\bibfield  {journal} {\bibinfo
  {journal} {Phys. Rev. B}\ }\textbf {\bibinfo {volume} {95}},\ \bibinfo
  {pages} {235107} (\bibinfo {year} {2017})}\BibitemShut {NoStop}%
\bibitem [{\citenamefont {{Jiang}}\ \emph {et~al.}(2016)\citenamefont
  {{Jiang}}, \citenamefont {{Kim}}, \citenamefont {{Han}},\ and\ \citenamefont
  {{Ran}}}]{jiang2016}%
  \BibitemOpen
  \bibfield  {author} {\bibinfo {author} {\bibfnamefont {S.}~\bibnamefont
  {{Jiang}}}, \bibinfo {author} {\bibfnamefont {P.}~\bibnamefont {{Kim}}},
  \bibinfo {author} {\bibfnamefont {J.~H.}\ \bibnamefont {{Han}}}, \ and\
  \bibinfo {author} {\bibfnamefont {Y.}~\bibnamefont {{Ran}}},\ }\bibfield
  {title} {\enquote {\bibinfo {title} {{Competing Spin Liquid Phases in the
  S=$\frac{1}{2}$ Heisenberg Model on the Kagome Lattice}},}\ }\href@noop {}
  {\bibfield  {journal} {\bibinfo  {journal} {ArXiv e-prints}\ } (\bibinfo
  {year} {2016})},\ \Eprint {http://arxiv.org/abs/1610.02024} {arXiv:1610.02024
  [cond-mat.str-el]} \BibitemShut {NoStop}%
\bibitem [{\citenamefont {He}\ \emph {et~al.}(2017)\citenamefont {He},
  \citenamefont {Zaletel}, \citenamefont {Oshikawa},\ and\ \citenamefont
  {Pollmann}}]{he2017}%
  \BibitemOpen
  \bibfield  {author} {\bibinfo {author} {\bibfnamefont {Yin-Chen}\
  \bibnamefont {He}}, \bibinfo {author} {\bibfnamefont {Michael~P.}\
  \bibnamefont {Zaletel}}, \bibinfo {author} {\bibfnamefont {Masaki}\
  \bibnamefont {Oshikawa}}, \ and\ \bibinfo {author} {\bibfnamefont {Frank}\
  \bibnamefont {Pollmann}},\ }\bibfield  {title} {\enquote {\bibinfo {title}
  {Signatures of dirac cones in a dmrg study of the kagome heisenberg model},}\
  }\href {\doibase 10.1103/PhysRevX.7.031020} {\bibfield  {journal} {\bibinfo
  {journal} {Phys. Rev. X}\ }\textbf {\bibinfo {volume} {7}},\ \bibinfo {pages}
  {031020} (\bibinfo {year} {2017})}\BibitemShut {NoStop}%
\bibitem [{\citenamefont {Liao}\ \emph {et~al.}(2017)\citenamefont {Liao},
  \citenamefont {Xie}, \citenamefont {Chen}, \citenamefont {Liu}, \citenamefont
  {Xie}, \citenamefont {Huang}, \citenamefont {Normand},\ and\ \citenamefont
  {Xiang}}]{liao2017}%
  \BibitemOpen
  \bibfield  {author} {\bibinfo {author} {\bibfnamefont {H.~J.}\ \bibnamefont
  {Liao}}, \bibinfo {author} {\bibfnamefont {Z.~Y.}\ \bibnamefont {Xie}},
  \bibinfo {author} {\bibfnamefont {J.}~\bibnamefont {Chen}}, \bibinfo {author}
  {\bibfnamefont {Z.~Y.}\ \bibnamefont {Liu}}, \bibinfo {author} {\bibfnamefont
  {H.~D.}\ \bibnamefont {Xie}}, \bibinfo {author} {\bibfnamefont {R.~Z.}\
  \bibnamefont {Huang}}, \bibinfo {author} {\bibfnamefont {B.}~\bibnamefont
  {Normand}}, \ and\ \bibinfo {author} {\bibfnamefont {T.}~\bibnamefont
  {Xiang}},\ }\bibfield  {title} {\enquote {\bibinfo {title} {Gapless
  spin-liquid ground state in the $s=1/2$ kagome antiferromagnet},}\ }\href
  {\doibase 10.1103/PhysRevLett.118.137202} {\bibfield  {journal} {\bibinfo
  {journal} {Phys. Rev. Lett.}\ }\textbf {\bibinfo {volume} {118}},\ \bibinfo
  {pages} {137202} (\bibinfo {year} {2017})}\BibitemShut {NoStop}%
\bibitem [{\citenamefont {Messio}\ \emph {et~al.}(2010)\citenamefont {Messio},
  \citenamefont {C\'epas},\ and\ \citenamefont {Lhuillier}}]{messio2010}%
  \BibitemOpen
  \bibfield  {author} {\bibinfo {author} {\bibfnamefont {L.}~\bibnamefont
  {Messio}}, \bibinfo {author} {\bibfnamefont {O.}~\bibnamefont {C\'epas}}, \
  and\ \bibinfo {author} {\bibfnamefont {C.}~\bibnamefont {Lhuillier}},\
  }\bibfield  {title} {\enquote {\bibinfo {title} {Schwinger-boson approach to
  the kagome antiferromagnet with dzyaloshinskii-moriya interactions: Phase
  diagram and dynamical structure factors},}\ }\href {\doibase
  10.1103/PhysRevB.81.064428} {\bibfield  {journal} {\bibinfo  {journal} {Phys.
  Rev. B}\ }\textbf {\bibinfo {volume} {81}},\ \bibinfo {pages} {064428}
  (\bibinfo {year} {2010})}\BibitemShut {NoStop}%
\bibitem [{\citenamefont {Dodds}\ \emph {et~al.}(2013)\citenamefont {Dodds},
  \citenamefont {Bhattacharjee},\ and\ \citenamefont {Kim}}]{dodds2013}%
  \BibitemOpen
  \bibfield  {author} {\bibinfo {author} {\bibfnamefont {Tyler}\ \bibnamefont
  {Dodds}}, \bibinfo {author} {\bibfnamefont {Subhro}\ \bibnamefont
  {Bhattacharjee}}, \ and\ \bibinfo {author} {\bibfnamefont {Yong~Baek}\
  \bibnamefont {Kim}},\ }\bibfield  {title} {\enquote {\bibinfo {title}
  {Quantum spin liquids in the absence of spin-rotation symmetry: Application
  to herbertsmithite},}\ }\href {\doibase 10.1103/PhysRevB.88.224413}
  {\bibfield  {journal} {\bibinfo  {journal} {Phys. Rev. B}\ }\textbf {\bibinfo
  {volume} {88}},\ \bibinfo {pages} {224413} (\bibinfo {year}
  {2013})}\BibitemShut {NoStop}%
\bibitem [{\citenamefont {Punk}\ \emph {et~al.}(2014)\citenamefont {Punk},
  \citenamefont {Chowdhury},\ and\ \citenamefont {Sachdev}}]{punk2014}%
  \BibitemOpen
  \bibfield  {author} {\bibinfo {author} {\bibfnamefont {Matthias}\
  \bibnamefont {Punk}}, \bibinfo {author} {\bibfnamefont {Debanjan}\
  \bibnamefont {Chowdhury}}, \ and\ \bibinfo {author} {\bibfnamefont {Subir}\
  \bibnamefont {Sachdev}},\ }\bibfield  {title} {\enquote {\bibinfo {title}
  {Topological excitations and the dynamic structure factor of spin liquids on
  the kagome lattice},}\ }\href {http://dx.doi.org/10.1038/nphys2887}
  {\bibfield  {journal} {\bibinfo  {journal} {Nature Physics}\ }\textbf
  {\bibinfo {volume} {10}},\ \bibinfo {pages} {289} (\bibinfo {year}
  {2014})}\BibitemShut {NoStop}%
\bibitem [{\citenamefont {Halimeh}\ and\ \citenamefont
  {Punk}(2016)}]{punk2016}%
  \BibitemOpen
  \bibfield  {author} {\bibinfo {author} {\bibfnamefont {Jad~C.}\ \bibnamefont
  {Halimeh}}\ and\ \bibinfo {author} {\bibfnamefont {Matthias}\ \bibnamefont
  {Punk}},\ }\bibfield  {title} {\enquote {\bibinfo {title} {Spin structure
  factors of chiral quantum spin liquids on the kagome lattice},}\ }\href
  {\doibase 10.1103/PhysRevB.94.104413} {\bibfield  {journal} {\bibinfo
  {journal} {Phys. Rev. B}\ }\textbf {\bibinfo {volume} {94}},\ \bibinfo
  {pages} {104413} (\bibinfo {year} {2016})}\BibitemShut {NoStop}%
\bibitem [{\citenamefont {Sherman}\ and\ \citenamefont
  {Singh}(2018)}]{sherman2018}%
  \BibitemOpen
  \bibfield  {author} {\bibinfo {author} {\bibfnamefont {Nicholas~E.}\
  \bibnamefont {Sherman}}\ and\ \bibinfo {author} {\bibfnamefont {Rajiv R.~P.}\
  \bibnamefont {Singh}},\ }\bibfield  {title} {\enquote {\bibinfo {title}
  {Structure factors of the kagome-lattice heisenberg antiferromagnets at
  finite temperatures},}\ }\href {\doibase 10.1103/PhysRevB.97.014423}
  {\bibfield  {journal} {\bibinfo  {journal} {Phys. Rev. B}\ }\textbf {\bibinfo
  {volume} {97}},\ \bibinfo {pages} {014423} (\bibinfo {year}
  {2018})}\BibitemShut {NoStop}%
\bibitem [{\citenamefont {{Laeuchli}}\ and\ \citenamefont
  {{Lhuillier}}(2009)}]{lauchli2009}%
  \BibitemOpen
  \bibfield  {author} {\bibinfo {author} {\bibfnamefont {A.}~\bibnamefont
  {{Laeuchli}}}\ and\ \bibinfo {author} {\bibfnamefont {C.}~\bibnamefont
  {{Lhuillier}}},\ }\bibfield  {title} {\enquote {\bibinfo {title} {{Dynamical
  Correlations of the Kagome S=1/2 Heisenberg Quantum Antiferromagnet}},}\
  }\href@noop {} {\bibfield  {journal} {\bibinfo  {journal} {ArXiv e-prints}\ }
  (\bibinfo {year} {2009})},\ \Eprint {http://arxiv.org/abs/0901.1065}
  {arXiv:0901.1065 [cond-mat.str-el]} \BibitemShut {NoStop}%
\bibitem [{\citenamefont {Gong}\ \emph {et~al.}(2015)\citenamefont {Gong},
  \citenamefont {Zhu}, \citenamefont {Balents},\ and\ \citenamefont
  {Sheng}}]{gong2015}%
  \BibitemOpen
  \bibfield  {author} {\bibinfo {author} {\bibfnamefont {Shou-Shu}\
  \bibnamefont {Gong}}, \bibinfo {author} {\bibfnamefont {Wei}\ \bibnamefont
  {Zhu}}, \bibinfo {author} {\bibfnamefont {Leon}\ \bibnamefont {Balents}}, \
  and\ \bibinfo {author} {\bibfnamefont {D.~N.}\ \bibnamefont {Sheng}},\
  }\bibfield  {title} {\enquote {\bibinfo {title} {Global phase diagram of
  competing ordered and quantum spin-liquid phases on the kagome lattice},}\
  }\href {\doibase 10.1103/PhysRevB.91.075112} {\bibfield  {journal} {\bibinfo
  {journal} {Phys. Rev. B}\ }\textbf {\bibinfo {volume} {91}},\ \bibinfo
  {pages} {075112} (\bibinfo {year} {2015})}\BibitemShut {NoStop}%
\bibitem [{\citenamefont {White}(1992)}]{white1992}%
  \BibitemOpen
  \bibfield  {author} {\bibinfo {author} {\bibfnamefont {Steven~R.}\
  \bibnamefont {White}},\ }\bibfield  {title} {\enquote {\bibinfo {title}
  {Density matrix formulation for quantum renormalization groups},}\ }\href
  {\doibase 10.1103/PhysRevLett.69.2863} {\bibfield  {journal} {\bibinfo
  {journal} {Phys. Rev. Lett.}\ }\textbf {\bibinfo {volume} {69}},\ \bibinfo
  {pages} {2863--2866} (\bibinfo {year} {1992})}\BibitemShut {NoStop}%
\bibitem [{\citenamefont {K\"uhner}\ and\ \citenamefont
  {White}(1999)}]{white1999}%
  \BibitemOpen
  \bibfield  {author} {\bibinfo {author} {\bibfnamefont {Till~D.}\ \bibnamefont
  {K\"uhner}}\ and\ \bibinfo {author} {\bibfnamefont {Steven~R.}\ \bibnamefont
  {White}},\ }\bibfield  {title} {\enquote {\bibinfo {title} {Dynamical
  correlation functions using the density matrix renormalization group},}\
  }\href {\doibase 10.1103/PhysRevB.60.335} {\bibfield  {journal} {\bibinfo
  {journal} {Phys. Rev. B}\ }\textbf {\bibinfo {volume} {60}},\ \bibinfo
  {pages} {335--343} (\bibinfo {year} {1999})}\BibitemShut {NoStop}%
\bibitem [{\citenamefont {Jeckelmann}(2002)}]{jeck2002}%
  \BibitemOpen
  \bibfield  {author} {\bibinfo {author} {\bibfnamefont {Eric}\ \bibnamefont
  {Jeckelmann}},\ }\bibfield  {title} {\enquote {\bibinfo {title} {Dynamical
  density-matrix renormalization-group method},}\ }\href {\doibase
  10.1103/PhysRevB.66.045114} {\bibfield  {journal} {\bibinfo  {journal} {Phys.
  Rev. B}\ }\textbf {\bibinfo {volume} {66}},\ \bibinfo {pages} {045114}
  (\bibinfo {year} {2002})}\BibitemShut {NoStop}%
\bibitem [{SM()}]{SM}%
  \BibitemOpen
  \href@noop {} {}\bibinfo {note} {See Supplemental Material for more
  details.}\BibitemShut {Stop}%
\bibitem [{\citenamefont {Kolley}\ \emph {et~al.}(2015)\citenamefont {Kolley},
  \citenamefont {Depenbrock}, \citenamefont {McCulloch}, \citenamefont
  {Schollw\"ock},\ and\ \citenamefont {Alba}}]{Kolley2015}%
  \BibitemOpen
  \bibfield  {author} {\bibinfo {author} {\bibfnamefont {F.}~\bibnamefont
  {Kolley}}, \bibinfo {author} {\bibfnamefont {S.}~\bibnamefont {Depenbrock}},
  \bibinfo {author} {\bibfnamefont {I.~P.}\ \bibnamefont {McCulloch}}, \bibinfo
  {author} {\bibfnamefont {U.}~\bibnamefont {Schollw\"ock}}, \ and\ \bibinfo
  {author} {\bibfnamefont {V.}~\bibnamefont {Alba}},\ }\bibfield  {title}
  {\enquote {\bibinfo {title} {Phase diagram of the ${J}_{1}\text{-}{J}_{2}$
  heisenberg model on the kagome lattice},}\ }\href {\doibase
  10.1103/PhysRevB.91.104418} {\bibfield  {journal} {\bibinfo  {journal} {Phys.
  Rev. B}\ }\textbf {\bibinfo {volume} {91}},\ \bibinfo {pages} {104418}
  (\bibinfo {year} {2015})}\BibitemShut {NoStop}%
\bibitem [{\citenamefont {Kalmeyer}\ and\ \citenamefont
  {Laughlin}(1987)}]{kalmeyer1987}%
  \BibitemOpen
  \bibfield  {author} {\bibinfo {author} {\bibfnamefont {V.}~\bibnamefont
  {Kalmeyer}}\ and\ \bibinfo {author} {\bibfnamefont {R.~B.}\ \bibnamefont
  {Laughlin}},\ }\bibfield  {title} {\enquote {\bibinfo {title} {Equivalence of
  the resonating-valence-bond and fractional quantum hall states},}\ }\href
  {\doibase 10.1103/PhysRevLett.59.2095} {\bibfield  {journal} {\bibinfo
  {journal} {Phys. Rev. Lett.}\ }\textbf {\bibinfo {volume} {59}},\ \bibinfo
  {pages} {2095--2098} (\bibinfo {year} {1987})}\BibitemShut {NoStop}%
\bibitem [{\citenamefont {Mikeska}\ and\ \citenamefont
  {Kolezhuk}(2004)}]{mikeska2004}%
  \BibitemOpen
  \bibfield  {author} {\bibinfo {author} {\bibfnamefont {Hans-J{\"u}rgen}\
  \bibnamefont {Mikeska}}\ and\ \bibinfo {author} {\bibfnamefont {Alexei~K}\
  \bibnamefont {Kolezhuk}},\ }\bibfield  {title} {\enquote {\bibinfo {title}
  {One-dimensional magnetism},}\ }\bibfield  {booktitle} {\emph {\bibinfo
  {booktitle} {Quantum magnetism}},\ }\href@noop {} {\  (\bibinfo {year}
  {2004})}\BibitemShut {NoStop}%
\bibitem [{\citenamefont {Moriya}(1960)}]{moriya1960}%
  \BibitemOpen
  \bibfield  {author} {\bibinfo {author} {\bibfnamefont {T\^oru}\ \bibnamefont
  {Moriya}},\ }\bibfield  {title} {\enquote {\bibinfo {title} {Anisotropic
  superexchange interaction and weak ferromagnetism},}\ }\href {\doibase
  10.1103/PhysRev.120.91} {\bibfield  {journal} {\bibinfo  {journal} {Phys.
  Rev.}\ }\textbf {\bibinfo {volume} {120}},\ \bibinfo {pages} {91--98}
  (\bibinfo {year} {1960})}\BibitemShut {NoStop}%
\bibitem [{\citenamefont {Zorko}\ \emph {et~al.}(2008)\citenamefont {Zorko},
  \citenamefont {Nellutla}, \citenamefont {van Tol}, \citenamefont {Brunel},
  \citenamefont {Bert}, \citenamefont {Duc}, \citenamefont {Trombe},
  \citenamefont {de~Vries}, \citenamefont {Harrison},\ and\ \citenamefont
  {Mendels}}]{zorko2008}%
  \BibitemOpen
  \bibfield  {author} {\bibinfo {author} {\bibfnamefont {A.}~\bibnamefont
  {Zorko}}, \bibinfo {author} {\bibfnamefont {S.}~\bibnamefont {Nellutla}},
  \bibinfo {author} {\bibfnamefont {J.}~\bibnamefont {van Tol}}, \bibinfo
  {author} {\bibfnamefont {L.~C.}\ \bibnamefont {Brunel}}, \bibinfo {author}
  {\bibfnamefont {F.}~\bibnamefont {Bert}}, \bibinfo {author} {\bibfnamefont
  {F.}~\bibnamefont {Duc}}, \bibinfo {author} {\bibfnamefont {J.-C.}\
  \bibnamefont {Trombe}}, \bibinfo {author} {\bibfnamefont {M.~A.}\
  \bibnamefont {de~Vries}}, \bibinfo {author} {\bibfnamefont {A.}~\bibnamefont
  {Harrison}}, \ and\ \bibinfo {author} {\bibfnamefont {P.}~\bibnamefont
  {Mendels}},\ }\bibfield  {title} {\enquote {\bibinfo {title}
  {Dzyaloshinsky-moriya anisotropy in the spin-1/2 kagome compound
  ${\mathrm{zncu}}_{3}(\mathrm{OH}{)}_{6}{\mathrm{cl}}_{2}$},}\ }\href
  {\doibase 10.1103/PhysRevLett.101.026405} {\bibfield  {journal} {\bibinfo
  {journal} {Phys. Rev. Lett.}\ }\textbf {\bibinfo {volume} {101}},\ \bibinfo
  {pages} {026405} (\bibinfo {year} {2008})}\BibitemShut {NoStop}%
\bibitem [{\citenamefont {Han}\ \emph {et~al.}(2012)\citenamefont {Han},
  \citenamefont {Chu},\ and\ \citenamefont {Lee}}]{han2012b}%
  \BibitemOpen
  \bibfield  {author} {\bibinfo {author} {\bibfnamefont {Tianheng}\
  \bibnamefont {Han}}, \bibinfo {author} {\bibfnamefont {Shaoyan}\ \bibnamefont
  {Chu}}, \ and\ \bibinfo {author} {\bibfnamefont {Young~S.}\ \bibnamefont
  {Lee}},\ }\bibfield  {title} {\enquote {\bibinfo {title} {Refining the spin
  hamiltonian in the spin-$\frac{1}{2}$ kagome lattice antiferromagnet
  ${\mathrm{zncu}}_{3}(\mathrm{OH}{)}_{6}{\mathrm{cl}}_{2}$ using single
  crystals},}\ }\href {\doibase 10.1103/PhysRevLett.108.157202} {\bibfield
  {journal} {\bibinfo  {journal} {Phys. Rev. Lett.}\ }\textbf {\bibinfo
  {volume} {108}},\ \bibinfo {pages} {157202} (\bibinfo {year}
  {2012})}\BibitemShut {NoStop}%
\bibitem [{\citenamefont {C\'epas}\ \emph {et~al.}(2008)\citenamefont
  {C\'epas}, \citenamefont {Fong}, \citenamefont {Leung},\ and\ \citenamefont
  {Lhuillier}}]{cepas2008}%
  \BibitemOpen
  \bibfield  {author} {\bibinfo {author} {\bibfnamefont {O.}~\bibnamefont
  {C\'epas}}, \bibinfo {author} {\bibfnamefont {C.~M.}\ \bibnamefont {Fong}},
  \bibinfo {author} {\bibfnamefont {P.~W.}\ \bibnamefont {Leung}}, \ and\
  \bibinfo {author} {\bibfnamefont {C.}~\bibnamefont {Lhuillier}},\ }\bibfield
  {title} {\enquote {\bibinfo {title} {Quantum phase transition induced by
  dzyaloshinskii-moriya interactions in the kagome antiferromagnet},}\ }\href
  {\doibase 10.1103/PhysRevB.78.140405} {\bibfield  {journal} {\bibinfo
  {journal} {Phys. Rev. B}\ }\textbf {\bibinfo {volume} {78}},\ \bibinfo
  {pages} {140405} (\bibinfo {year} {2008})}\BibitemShut {NoStop}%
\bibitem [{\citenamefont {Huh}\ \emph {et~al.}(2010)\citenamefont {Huh},
  \citenamefont {Fritz},\ and\ \citenamefont {Sachdev}}]{huh2010}%
  \BibitemOpen
  \bibfield  {author} {\bibinfo {author} {\bibfnamefont {Yejin}\ \bibnamefont
  {Huh}}, \bibinfo {author} {\bibfnamefont {Lars}\ \bibnamefont {Fritz}}, \
  and\ \bibinfo {author} {\bibfnamefont {Subir}\ \bibnamefont {Sachdev}},\
  }\bibfield  {title} {\enquote {\bibinfo {title} {Quantum criticality of the
  kagome antiferromagnet with dzyaloshinskii-moriya interactions},}\ }\href
  {\doibase 10.1103/PhysRevB.81.144432} {\bibfield  {journal} {\bibinfo
  {journal} {Phys. Rev. B}\ }\textbf {\bibinfo {volume} {81}},\ \bibinfo
  {pages} {144432} (\bibinfo {year} {2010})}\BibitemShut {NoStop}%
\bibitem [{\citenamefont {{Sun}}\ \emph {et~al.}(2018)\citenamefont {{Sun}},
  \citenamefont {{Wang}}, \citenamefont {{Fang}}, \citenamefont {{Qi}},
  \citenamefont {{Cheng}},\ and\ \citenamefont {{Meng}}}]{Sun2018}%
  \BibitemOpen
  \bibfield  {author} {\bibinfo {author} {\bibfnamefont {G.~Y.}\ \bibnamefont
  {{Sun}}}, \bibinfo {author} {\bibfnamefont {Y.-C.}\ \bibnamefont {{Wang}}},
  \bibinfo {author} {\bibfnamefont {C.}~\bibnamefont {{Fang}}}, \bibinfo
  {author} {\bibfnamefont {Y.}~\bibnamefont {{Qi}}}, \bibinfo {author}
  {\bibfnamefont {M.}~\bibnamefont {{Cheng}}}, \ and\ \bibinfo {author}
  {\bibfnamefont {Z.~Y.}\ \bibnamefont {{Meng}}},\ }\bibfield  {title}
  {\enquote {\bibinfo {title} {{Dynamical Signature of Symmetry
  Fractionalization in Frustrated Magnets}},}\ }\href@noop {} {\bibfield
  {journal} {\bibinfo  {journal} {ArXiv e-prints}\ } (\bibinfo {year}
  {2018})},\ \Eprint {http://arxiv.org/abs/1803.10969} {arXiv:1803.10969
  [cond-mat.str-el]} \BibitemShut {NoStop}%
\bibitem [{\citenamefont {{Becker}}\ and\ \citenamefont
  {{Wessel}}(2018)}]{Becker2018}%
  \BibitemOpen
  \bibfield  {author} {\bibinfo {author} {\bibfnamefont {J.}~\bibnamefont
  {{Becker}}}\ and\ \bibinfo {author} {\bibfnamefont {S.}~\bibnamefont
  {{Wessel}}},\ }\bibfield  {title} {\enquote {\bibinfo {title} {{Diagnosing
  Fractionalization from the Spin Dynamics of $Z\_2$ Spin Liquids on the Kagome
  Lattice by Quantum Monte Carlo}},}\ }\href@noop {} {\bibfield  {journal}
  {\bibinfo  {journal} {ArXiv e-prints}\ } (\bibinfo {year} {2018})},\ \Eprint
  {http://arxiv.org/abs/1803.10970} {arXiv:1803.10970 [cond-mat.str-el]}
  \BibitemShut {NoStop}%
\bibitem [{\citenamefont {Hastings}(2000)}]{Hastings2000}%
  \BibitemOpen
  \bibfield  {author} {\bibinfo {author} {\bibfnamefont {M.~B.}\ \bibnamefont
  {Hastings}},\ }\bibfield  {title} {\enquote {\bibinfo {title} {Dirac
  structure, rvb, and goldstone modes in the kagome antiferromagnet},}\ }\href
  {\doibase 10.1103/PhysRevB.63.014413} {\bibfield  {journal} {\bibinfo
  {journal} {Phys. Rev. B}\ }\textbf {\bibinfo {volume} {63}},\ \bibinfo
  {pages} {014413} (\bibinfo {year} {2000})}\BibitemShut {NoStop}%
\bibitem [{\citenamefont {Wang}\ and\ \citenamefont
  {Vishwanath}(2006)}]{WangFa2006}%
  \BibitemOpen
  \bibfield  {author} {\bibinfo {author} {\bibfnamefont {Fa}~\bibnamefont
  {Wang}}\ and\ \bibinfo {author} {\bibfnamefont {Ashvin}\ \bibnamefont
  {Vishwanath}},\ }\bibfield  {title} {\enquote {\bibinfo {title} {Spin-liquid
  states on the triangular and kagome lattices: A projective-symmetry-group
  analysis of schwinger boson states},}\ }\href {\doibase
  10.1103/PhysRevB.74.174423} {\bibfield  {journal} {\bibinfo  {journal} {Phys.
  Rev. B}\ }\textbf {\bibinfo {volume} {74}},\ \bibinfo {pages} {174423}
  (\bibinfo {year} {2006})}\BibitemShut {NoStop}%
\bibitem [{\citenamefont {Laughlin}(1983)}]{Laughlin1983}%
  \BibitemOpen
  \bibfield  {author} {\bibinfo {author} {\bibfnamefont {R.~B.}\ \bibnamefont
  {Laughlin}},\ }\bibfield  {title} {\enquote {\bibinfo {title} {Anomalous
  quantum hall effect: An incompressible quantum fluid with fractionally
  charged excitations},}\ }\href {\doibase 10.1103/PhysRevLett.50.1395}
  {\bibfield  {journal} {\bibinfo  {journal} {Phys. Rev. Lett.}\ }\textbf
  {\bibinfo {volume} {50}},\ \bibinfo {pages} {1395--1398} (\bibinfo {year}
  {1983})}\BibitemShut {NoStop}%
\end{thebibliography}%

\clearpage
\begin{widetext}
\widetext

\appendix
\begin{appendices}

\section{Numerical Method}\label{app:dmrg}
In this section, we introduce the numerical simulation details about dynamical properties in
density-matrix renormalization group (DMRG) algorithm. We also provide a benchmark on square
Heisenberg model to show the high accuracy of DMRG algorithm.

\subsection{1. Density-matrix renormalization group algorithm}
We perform the calculations based on high accuracy DMRG
on cylinder geometry with closed boundary in the y direction
and open boundary in the x direction. We denote it as
$L_y\times L_x$ ($L_x\gg L_y$), where $L_y$ and $L_x$ are the number of unit cells in the
y and x directions.
We first perform the ground state  DMRG procedure and sweep the ground state on the whole cylinder,
and then target the dynamical properties (see below) by sweeping the
middle $L_y \times L_y$ unit cells to avoid edge excitations.
Most of the calculations are performed on $L_y=4$ and $L_y=6$ cylinders.

Here we would like to point out that, although the whole system on cylinder does not host
translational symmetry along x-direction, the ground state in the middle of a long cylinder
approximatelly satisfies the translational symmetry (the emergent translational period determined by the nature of the ground state itself).
Due to this reason, we can cut the middle $L_y\times L_y$ system and glue it into a torus (with periodic boundary condition along both x- and y-direction),
so that the momentum quantum number can be well defined along both x- and y-direction (within $L_y\times L_y$ unit cells in the middle of the cylinder).
This process is widely used for spin structure factor calculations in DMRG community.

The conventional DMRG algorithm only targets the ground state, $|0\rangle$.
To calculate dynamical spin structure factor, we apply the dynamical DMRG
by targeting the following states together with the ground state when sweeping:
\begin{center}
\begin{tabular}{lll}
		$|S^\alpha(\mathbf Q) \rangle$ & $= S^\alpha(\mathbf Q) \,| 0 \rangle$ & 	\\
		$|x^\alpha(\omega+i\eta) \rangle$ &$=
	\frac{1}{\omega+i\eta-(H-E_{GS})} \, {|S^\alpha(\mathbf Q) \rangle}$ &
\end{tabular}
\end{center}
where $|x(\omega) \rangle$  is usually called \textit{correction vector}
which can be calculated by the conjugate gradient method~\cite{white1999} or other algorithm \cite{jeck2002}.
With the help of the correction vector, the dynamical spin structure factor can be calculated directly:
\begin{equation}
	\mathcal{S}^{\alpha\beta}(\mathbf{Q},\omega) = -\frac{1}{\pi}Im\langle S^\alpha(\mathbf Q)|  x^\beta(\omega+i\eta) \rangle
\end{equation}
where $\eta$ takes  a small  positive value as the smearing energy.
Taking these states ($|0\rangle$,$|S^\alpha(\mathbf Q)\rangle$ and $|x^\alpha(\omega+i\eta) \rangle$)
as target states and optimizing the DMRG basis to represent them allow for
a precise calculation of the structure factor for
a given frequency $\omega$ and the broadening factor $\eta$.
In this work, all calculations are performed using $\eta=0.05$ and $\eta=0.1$ (in unit of nearest-neighbor coupling $J_1$).
Since we have to target multi-states in the DMRG process,
the truncation error is basically larger than the ground state  DMRG.
In this work, we ensure the truncation error of the order or smaller than $10^{-5}$,
by keeping up to $2400$ states.

Here we also comment on the numerical scheme we used in this paper. In general, there are two main algorithms  to
target dynamics based on DMRG algorithm. One is to calculate the dynamical spin structure factor in the frequency regime (as outlined above), the other one
is to first calculate the time-evolution of the physical quantities and then obtain the dynamical spin structure factor by Fourier transformation.
In general, the first method is more accurate in the low-frequency regime, while the second method works better in high-frequency regime
(because the accumulated errors grow as time steps increases in time-dependent DMRG).
Based on this reason, in the discussion of low-energy physics of kagome Heisenberg model, we utilize
the first method (\cite{white1999,jeck2002}) in this paper.

\subsection{2. A benchmark: Neel order on square lattice}
In this section, as a benchmark of DMRG method,
we show the dynamical spin structure factor of the $S = 1/2$ antiferromagnetic Heisenberg model on the square lattice.
For $J_1$ Heisenberg model on square lattice, the ground state is  a $q=(\pi,\pi)$ Neel ordered state.
Thus, we  expect to see  a single-mode gapless excitation dispersion related to magnon quasiparticle in dynamical spin structure factor \cite{Shimokawa2015}.
 Fig. \ref{fig:dynstat_sq_1} shows  the energy-dependence of the dynamical spin structure factor
$S(\mathbf Q, \omega)$ at several typical momenta.
We further extract the peak position at each momentum point and plot the single magnon dispersion in
Fig. \ref{fig:dynstat_sq_2}, which is in largely agreement with the spin wave theory
(For $q=(\pi,\pi)$ Neel order, it is believed that spin wave theory can capture the main features of dynamics except for $\mathbf Q=M$ point.).
The dominant peak is a single magnon excitation.
Importantly, the largest weight is carried by $S(\mathbf Q=X,\omega=0)$,
and the peak location of $\mathbf Q=X=(\pi,\pi)$ centered at $\omega=0$ directly reveals the gapless nature for Neel $q=(\pi,\pi)$ phase.
Interestingly, the $S(\mathbf Q=M, \omega)$ shows an anomalous tail in high energy regime, which could be attributed to the fact
that magnon-magnon interaction is enhanced near $\mathbf Q=M$ \cite{Powalski2017}.
The discrepancy  near the $M$ point  between our result and the linear spin wave theory comes from that
a linear spin wave theory neglects the magnon-magnon interactions (with first-order corrections,
it is known that single magnon dispersion shifts upward).
Here, through this benchmark on the square lattice and extensive tests on one dimensional chain (not shown here), we conclude that the
current scheme can obtain  reliable dynamical properties efficiently.
Next we will apply the above strategy to antiferromagnetic Heisenberg model on the kagome lattice.

\begin{figure*}[!htb]
\includegraphics[width=0.35\linewidth]{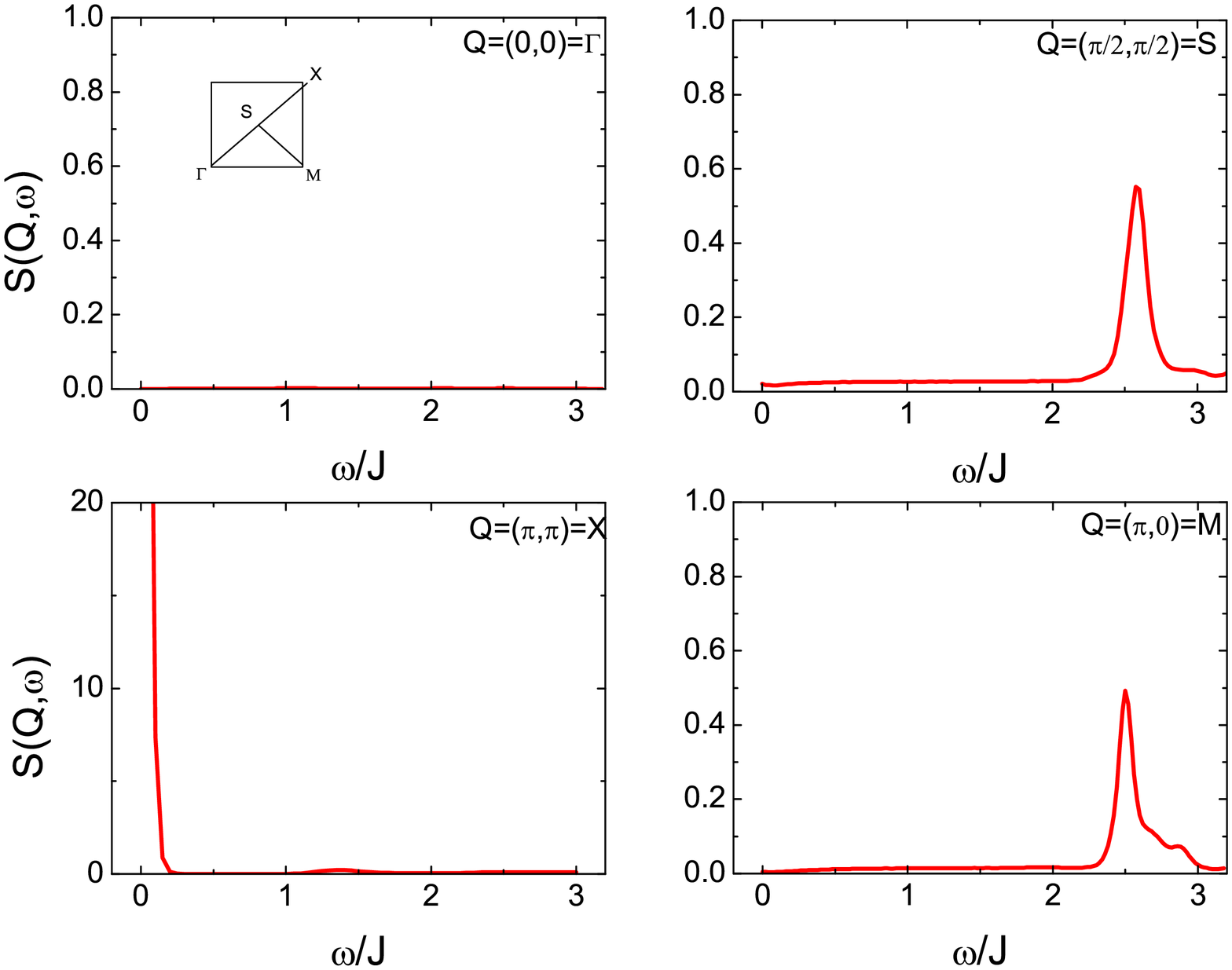}
\caption{The energy scan of dynamical spin structural factor $\mathcal S(Q,\omega)$ for Heisenberg model on the square lattice,
for several typical momentum points in BZ. The calculations are performed on $L_y=8$ cylinder by keeping $M=400$ states.
\label{fig:dynstat_sq_1}}
\end{figure*}

\begin{figure*}[!htb]
\includegraphics[width=0.4\linewidth]{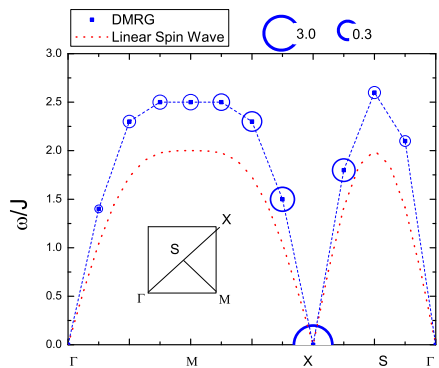}
\caption{Synthetic the peak position (blue dots) of the longitudinal dynamical structure factors along a path of highly symmetric points in the Brillouin
zone. The size of each blue circle is proportional to the static spin structure factor by sum up dynamical spin structure factor over energy.
Red solid line shows the dispersions obtained within linear (harmonic) spin-wave theory.
\label{fig:dynstat_sq_2}}
\end{figure*}

\subsection{3. Analysis of the Finite-size Effect}
In this work, we utilize the DMRG algorithm to simulate the dynamical response.
Although we can easily go beyond the exact diagonalization limit,
the DMRG calculation still suffers from the finite-size effect, which will be discussed in detail here.

Since the cylinder geometry is preferred in DMRG algorithm,
the available lattice system sizes are  limited by the width along  the wrapped direction similar to the ground state DMRG
(saying, $L_y$, which accounts the number of unit cells in the wrapped direction).
On the kagome lattice, the current computational ability is limited to accessing $L_y$ up to $6$,
depending on the nature of different phases. To be specific,
the largest system size is $L_y=6$ for Neel $q=(0,0)$ order and chiral spin liquid.
While for quantum spin liquid the largest available system is $L_y=4$, because
the highly frustrated nature near $J_1$ Heisenberg point leads to the slow convergence in dmrg calculations.

We have extensively checked that, the main features of dynamical spin structure factor are robust
for Neel $q=(0,0)$ phase and chiral spin liquid phase, by tuning the system sizes $L_y=4,5,6$.
Thus, we have confidence that the nature of dynamical responses of these two phases shown in the main text is intrinsic properties
of the corresponding two-dimensional systems.

Nevertheless, for the quantum spin liquid phase, we cannot fully rule out the finite-size effect based on $L_y=4$ system,
since $L_y=4$ is the only available system size. (We cannot reach a converged ground state in
dynamical dmrg algorithm for $L_y=5,6$ for quantum spin liquid phase due to the difficulty in convergence in such a state).
In the main text, we utilize the twisted boundary condition to inspect the gapless nature on a given finite-size system.
The main physical reason is further clarified here.
First, tuning the boundary condition is a general method to detect the nature of ground state on finite-size calculations.
Since the available discrete momentum vectors are limited due to the finite-size effect, tuning the twisted boundary condition
allows us to reach more momentum points in the Brilliuin zone.
Second,  for ground states with intrinsic topological orders, it is expected that
the ground state manifold is robust to the twisted boundary condition, without energy level crossing with higher energy levels.
In contrast, energy level crossing may occur by tuning boundary conditions if the ground state is gapless.
Here the picture is akin to the Thouless's picture of localization:
The energy spectral flow of insulators is robust against boundary conditions, however, energy flow of metallic phase is not.
For gapless phase, the change of energy spectrum by twisted boundary conditions
inevitably leads to substantial difference in dynamical response functions.
Based on these  reasons, we inspect the dynamical response for quantum spin liquid phase
by tuning different boundary conditions. This is a way out for uncovering the intrinsic nature of ground state on the finite-size calculation.

\begin{figure}[t]
\includegraphics[width=0.45\textwidth]{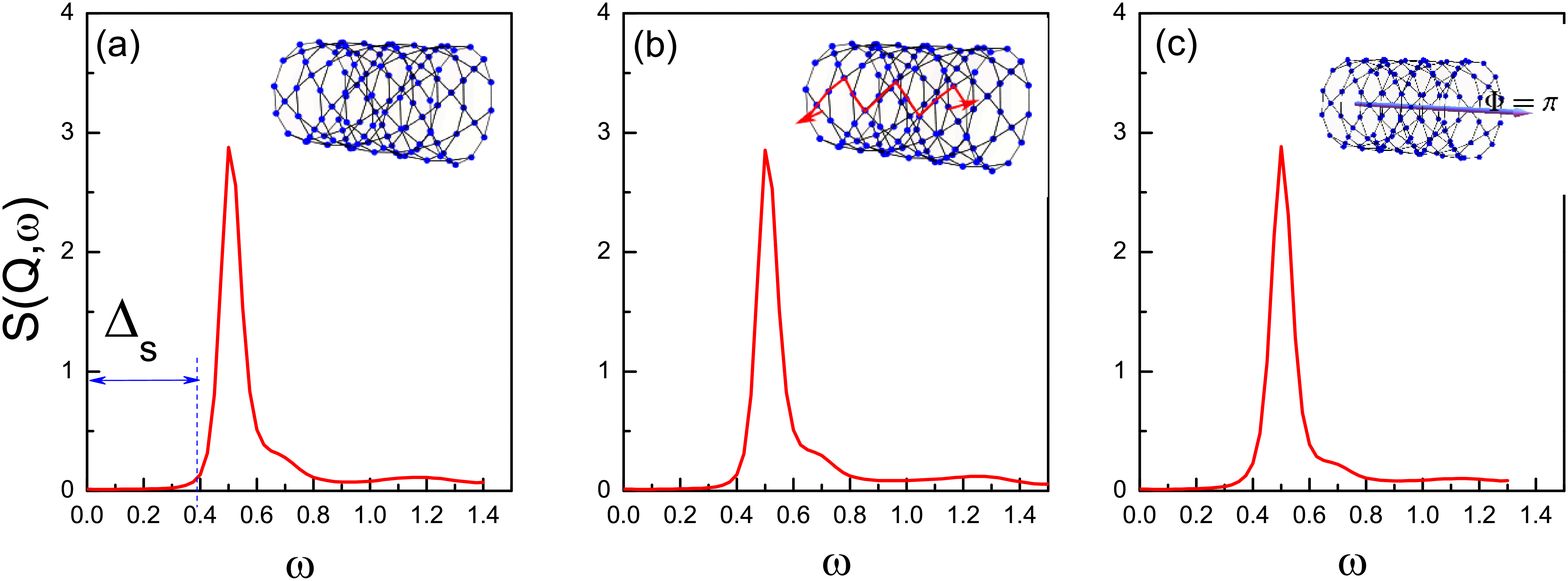}\\
\includegraphics[width=0.45\textwidth]{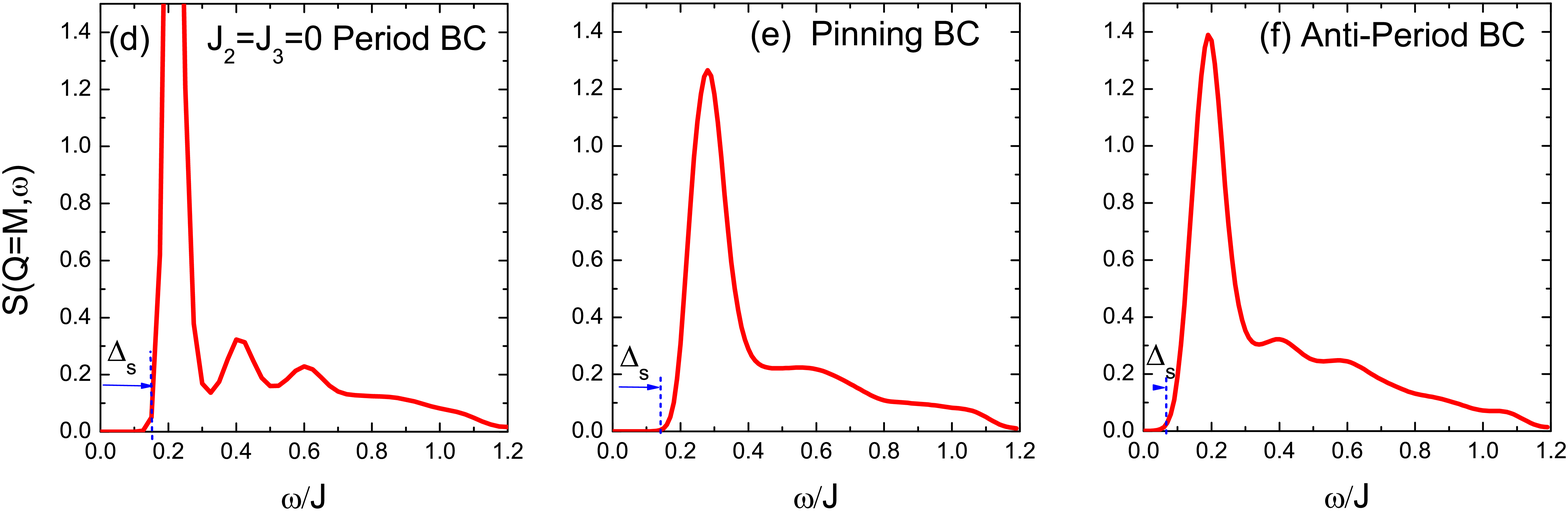}
\caption{DSSF for CSL phase (a-c) and KSL phase (d-f)
for different boundary conditions (BCs):
(a,d) periodic BC in wrapped direction (b,e) periodic BC on wrapped direction and a pinning field on the open direction,
and (c,f) anti-periodic BC in wrapped direction.
}\label{fig:dsq_kag_BC}
\end{figure}

\subsection{4. Tuning boundary conditions}
We address the question  whether or not the ground state of the KSL is gapped,
which holds the clue to distinguish the different theoretical scenarios \cite{Ran2007}\cite{Hastings2000}\cite{Sachdev1992}\cite{WangFa2006},
by inspecting the response of the system under   tuning different BCs.
As a benchmark, we first test the system in the gapped CSL phase.
Since the CSL is equivalent to the $\nu = 1/2$ bosonic Laughlin state~\cite{Laughlin1983},
the system has two-fold topological degenerate ground states. In DMRG simulation,
the ground state in the spinon sector can be obtained by adiabatically changing the BC
by a $2\pi$ phase. As shown in Fig.~\ref{fig:dsq_kag_BC}(a-b), the DSSF in the two
ground states are almost identical, which can be understood by the fact that local
measurements are unable to distinguish different topologically degenerate
ground states.

Now we inspect the response of the KSL.
Fig. \ref{fig:dsq_kag_BC}(d) shows $\mathcal S(\mathbf Q=M,\omega)$,
by imposing periodic BC on the wrapped direction (same with Fig. \ref{fig:dsq}).
As a comparison, Fig. \ref{fig:dsq_kag_BC}(e) shows the case of additionally pinning a spinon at each open end of cylinder geometry. 
Although the spin gap remains robust,
the predominant spectral peak in low-energy regime becomes broader.
Moreover, by imposing anti-periodic BC in wrapped direction, as shown in Fig. \ref{fig:dsq_kag_BC}(f),
the excitation gap $\Delta_s$ shrinks from $\Delta_s\approx0.16$ to a smaller value $\approx 0.075$, 
signaling that the spin excitation gap is sensitive to the BC.
Here, the dramatical change of lineshape of spectral peak and the shrink of spin gap in DSSF,
indicate that the ground state is near critical or having very small gap.
Of course,  the finite size effect is generally important,  
which calls for future work on finite-size scaling analysis.

\section{Transition from chiral spin liquid phase to Neel $q=(0,0)$ phase}\label{app:CSL-Neel}
In this section, we study the phase transition from the chiral spin liquid phase to magnetic Neel $q=(0,0)$ phase.
This phase transition is interesting due to the following reasons.
First, it is intriguing to understand the low-energy peak structure in dynamical spin structure factor at $\mathbf Q=M$ point.
Second, it is an exotic example of continuous phase transition between gapped topological ordered state and topological trivial state.

According to the global phase diagram in the main text (Fig. 1(a)), for finite $J_2>0.15$
different phases may appear depending on $J_3$.
Tuning $J_3$ will drive a phase transition from chiral spin liquid phase
to magnetic $q=(0,0)$ phase.
And it has been found that, chiral spin liquid undergoes a \textit{continuous} phase transition
to Neel $q=(0,0)$ phase \cite{gong2015},
as evidenced by the fact that all local order parameters change smoothly across the phase transition point.
However, the reason for this continuous phase transition is less understood before,
because the transition between a gapped topological ordered phase and a topological trivial phase is often  thought to be first-order type.

Next we will show the evolution of dynamical spin structure factor for various $J_3$, by setting $J_2=0.25 J_1$.
We will focus on momentum wave vector $\mathbf Q=M$ in this section.
As shown in Fig. \ref{sfig:CSL-Neel}, we show the evolution of dynamical spin structure factor at momentum point $\mathbf Q=M$,
for various $J_3$. The key features are:
1) In chiral spin liquid phase ($J_3>0.18 J_1$), there exists a peak structure at low frequency regime at $\mathbf Q=M$ point,
as discussed in the main text.
By approaching the transition point, this peak structure moves towards the low-frequency regime, and peak intensity
gradually increases.
2) In the Neel order phase, the peak position is centered at $\omega=0$.
3) In chiral spin liquid phase, the predominant peak is connected to the high-frequency spin continuum, while
in Neel phase the zero-frequency peak is well separated from high-frequency spin excitations.

Based on the above observations, a natural interpretation of the peak structure in chiral spin liquid phase is two-spinon resonance state.
The reason is that,
it is well-known the peak at $\omega=0$ in Neel phase relates to magnon quasiparticle, which can be viewed as a bound state of two spinons.
Taking into account that the element excitation in the chiral spin liquid phase is \textit{deconfined} spinon,
we can take the peak in chiral spin liquid as two-spinon resonance state, while
the peak in Neel phase as two-spinon bound state (equivalent to magnon state).
Two-spinon resonance naturally depends on the interaction coupling $J_3$.
By approaching critical point, two-spinon resonance moves towards zero frequency.
In the vicinity of the critical point, two-spinon resonance state becomes two-spinon bound state (equivalent to a magnon).
The further condensation of pair spinons  should lead to formation of Neel magnetic order.
In a word, this picture leads to two important physics:
First, the peak of dynamical spin structure factor at momentum $\mathbf Q=M$ in chiral spin liquid can be interpreted as two-spinon resonance state.
Second, the transition from chiral spin liquid to Neel phase can be understood by
the formation of condensate of two-spinon bound state or magnon. It therefore provides
a microscopic understanding of continuous phase transition between
chiral spin liquid and Neel phase.

In the above analysis, the peak structure in dynamic spin structure factor of Neel phase and chiral spin liquid occur at the same momentum point ($\mathbf Q=M$),
which makes the mechanism of spinon pair condensate possible.
If the magnetic wave vector of underlying long-ranged magnetic order is different from that of two-spinon resonance state in spin liquid,
the phase transition from chiral spin liquid to magnetic ordered phase should be first order.
For example, the transition from chiral spin liquid to cuboc1 phase in the global phase diagram is first-order type \cite{gong2015}.
To sum up, the evolution of dynamical spin structure factor acrossing the critical point,
not only elucidate the nature of the ground state,
but also provides invaluable insights on the nature of related phase transition.


\section{Spin correlations and spin gap in the presence of DM interaction}\label{app:DM}
In the main text, we show a phase diagram as a function of DM interaction $D^z$ and next-nearest-neighbor coupling $J_2$.
The phase boundary between spin liquid phase and magnetic $q=(0,0)$ phase is determined by the spin gap and spin correlations.
In Fig. \ref{sfig:DM_corr}(a-b), we show the spin gap dependence on parameter $J_2$ and $D^z$, respectively.
It is found that the spin gap decreases monotonically as approaching the phase boundary.
In particular, in the spin liquid phase, the spin gap strongly depends on the twisted boundary condition,
indicating the finite spin gap is due to finite-size effect.
In contrast, in the magnetic ordered phase, the spin gap has little dependence on twisted boundary condition.
As shown in Fig. \ref{sfig:DM_corr}(c), the spin correlation provides another evidence for phase boundary.
The spin correlation exponentially decays with the distance, for $D^z<0.08$ and $J_2=0.0$.
For $D^z\geq0.08$, the long-ranged order emerges as the spin correlation tends to saturate.
Based on these facts, we determine the $D^z\approx 0.08$ as the phase boundary at $J_2=0.0$,
which is largely consistent with the previous estimation from ED calculation \cite{Messio2010}.
Using the similar method, we determine the phase boundary for non-zero $J_2$ case,
and map out the full phase diagram as shown in the main text.

\begin{figure}[!htb]
\includegraphics[width=0.65\textwidth]{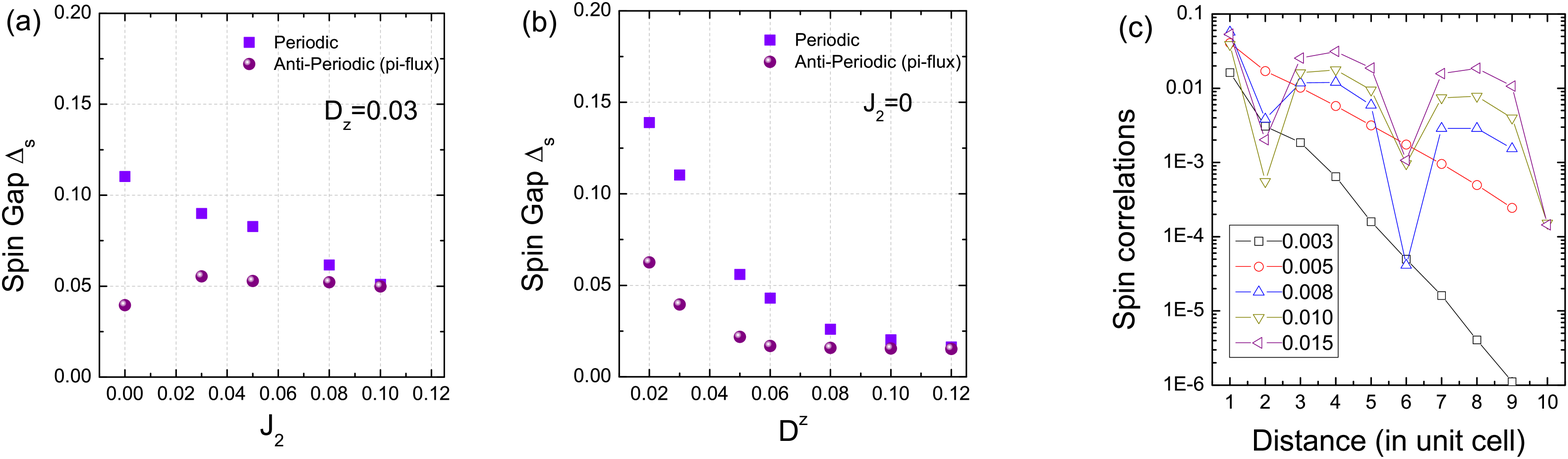}
\caption{(a) Spin gap as a function of $D^z$ by setting $J_2=0$ and (b) spin gap as a function of $J_2$ by setting $D^z=0$.
Spin gap is obtained by $\Delta_s=E_0(S^{tot}_z=1)-E_0(S^{tot}_z=0)$, where the lowest energy state of $S^{tot}_z=1$ is computed by targeting $S^{tot}_z=1$
in the center of the cylinder based on the ground state in $S^{tot}_z=0$.
(c) spin correlations $\langle S^+_i S^-_{i+d}\rangle$ for various $D^z$, by setting $J_2=0.0$.
These results are obtained on $L_y=4$ cylinder.
}\label{sfig:DM_corr}
\end{figure}

\end{appendices}

\end{widetext}

\end{document}